\documentclass[11pt, onecolumn, a4paper]{article}

\usepackage{amssymb, amsmath}
\usepackage{graphicx}  

\usepackage{algorithm}
\usepackage[noend]{algpseudocode}

\newcommand{\nop}[1]{}

\newtheorem{definition}{Definition}

\newtheorem{theorem}{Theorem}
\newtheorem{corollary}{Corollary}
\newtheorem{lemma}{Lemma}
\newtheorem{proposition}{Proposition}

\begin{document}

\title{Some Properties of Length Rate Quotient Shapers} 
\author{Yuming Jiang \\ NTNU -- Norwegian University of Science and Technology}      
\maketitle

\begin{abstract}
Length Rate Quotient (LRQ) is the first algorithm of interleaved shaping -- a novel concept proposed to provide per-flow shaping for a flow aggregate without per-flow queuing. This concept has been adopted by Time-Sensitive Networking (TSN) and Deterministic Networking (DetNet). An appealing property of interleaved shaping is that, when an interleaved shaper is appended to a FIFO system, it does not increase the worst-case delay of the system. Based on this ``shaping-for-free'' property, an approach has been introduced to deliver bounded end-to-end latency. Specifically, at each output link of a node, class-based aggregate scheduling is used together with one interleaved shaper per-input link and per-class, and the interleaved shaper re-shapes every flow to its initial traffic constraint. 
In this paper, we investigate other properties of interleaved LRQ shapers, particularly as stand-alone elements. 
In addition, under per-flow setting, we also investigate per-flow LRQ based flow aggregation and derive its properties. The analysis focuses directly on the timing of operations, such as shaping and scheduling, in the network. This timing based method can be found in the Guaranteed Rate (GR) server model and more generally the max-plus branch of network calculus. 
With the derived properties, we not only show that an improved end-to-end latency bound can be obtained for the current approach, but also demonstrate with two examples that new approaches may be devised.  
End-to-end delay bounds for the three approaches are derived and compared. As a highlight, the two new approaches do not require different node architectures in allocating (shaping / scheduling) queues,  which implies that they can be readily adapted for use in TSN and DetNet. This together with the derived properties of LRQ shed new insights on providing the TSN / DetNet qualities of service. 
\end{abstract}

{\bf Keywords:} Interleaved Shaping; Length Rate Quotient (LRQ); Time-Sensitive Networking (TSN); Deterministic Networking (DetNet); Asynchronous Traffic Shaping; Interleaved Shaper; Interleaved Regulator (IR)

\section{Introduction}\label{sec-1}

Interleaved shaping is a novel concept for traffic shaping, originally proposed by Specht and Samii in \cite{ECRTS16}. Conceptually, its idea is to perform per-flow traffic shaping within a flow aggregate using only one FIFO queue. An appealing property of interleaved shaping is the so-called ``shaping-for-free'' property: When an interleaved shaper is appended to a FIFO system and shapes flows to their initial traffic constraints, it does not increase the worst-case delay of the system. Based on this property, Specht and Samii also proposed in \cite{ECRTS16} an approach to achieve bounded worst-case end-to-end (e2e) delay in the network. The approach includes a specific way to allocate shaping and scheduling queues in switches and re-shaping flows to their initial traffic constraints using the corresponding interleaved shaping algorithms. 

The concept of interleaved shaping, together with the approach of allocating queues and reshaping traffic, has been adopted and extended by IEEE Time-Sensitive Networking (TSN) \cite{TSN} and IETF Deterministic Networking (DetNet) \cite{DetNet} to deliver bounded e2e latency. The concept is called Asynchronous Traffic Shaping (ATS) in the former \cite{Std-Qcr} while Interleaved Regulation in the latter \cite{Finn21}. 

In \cite{ECRTS16}, two algorithms for interleaved shaping are introduced, which are Length Rate Quotient (LRQ) and Token Bucket Emulation (TBE), together with a timing-based analysis on the worst-case e2e delay achieved by them. While LRQ is for traffic constraints where the gap between consecutive packets satisfies a length rate quotient condition, TBE is for the well-known token bucket (TB) or leaky bucket (TB) traffic constraints. In \cite{LeBoudec18}, more types of traffic constraints are investigated under a unified traffic constraint concept called ``Pi-regularity'' and the resultant interleaved shapers are called Interleaved Regulators (IRs). The ``shaping-for-free'' property is also proved for IRs in \cite{LeBoudec18}. 

Surprisingly, other than the ``shaping-for-free'' property, few other properties of interleaved shapers have been reported. As a step towards filling the gap, this paper is intended. Specifically, we focus on {\em LRQ, the first interleaved shaping algorithm} and derive its properties under both the interleaved setting and the per-flow setting. For interleaved LRQ shapers, in addition to ``shaping-for-free'', the proved properties include conformance, output characterization, a sufficient and necessary condition to ensure the existence of bounded delay, service characterization, and delay bounds. For per-flow LRQ shapers, in addition to the properties as a special case of the interleaved version, we particularly investigate properties of a per-flow LRQ based flow aggregator. 

Similar to the analysis in \cite{ECRTS16},  ours also employs timing based analysis, which directly investigates the timing of various operations, such as shaping and scheduling, in the network and the time relationships between them. Generally, this timing based analysis method can be found in the max-plus branch of network calculus (NC) \cite{Chang00} \cite{NetCal}. In this paper, for server modeling, rather than taking the (min-plus) service curve model \cite{NetCal} or the max-plus NC version of service curve, i.e., the $g$-server model \cite{Chang00}, we particularly adopt the Guaranteed Rate (GR) server model \cite{GLV95, GV97}, based on which the various properties are derived. An underlying motivation is the known fact that, without additional treatment, the directly obtained delay bound based on service curve models is looser than that from GR: See, e.g., \cite{Chang00} \cite{NetCal} for discussion about the treatment and \cite{Jiang-ITC13} for a timing analysis based discussion about the underlying reason and its impacts. 

With the derived properties, we discuss that an improved e2e delay bound can be obtained for the approach proposed in  \cite{ECRTS16} , in comparison with related bounds found in the literature, e.g., \cite{ECRTS16} for TSN ATS \cite{Std-Qcr}  and  \cite{ITC18} for DetNet \cite{Finn21}. This improvement is due to the adopted GR-based timing analysis. In \cite{ITC18} and indeed in most TSN / DetNet delay bound analysis literature as reviewed and discussed in \cite{Zhao21}, the analysis is based on the service curve server model. To illustrate the difference, strict priority is specifically used as an example, whose delay bounds, obtained using the service curve model \cite{ITC18}, the timing method in  \cite{ECRTS16} and the adopted GR model, are compared. 

In addition, we demonstrate with two examples that new approaches, based on the derived properties, may be devised which can also deliver bounded e2e latency. A comparison of the e2e delay bounds from the three approaches suggests that, by employing specific information of the network, the accordingly devised approaches may be able to offer better e2e delay bounds, in comparison to the universal approach \cite{ECRTS16}.

The rest of this paper is organized as follows. In the next section, i.e, Section \ref{sec-2}, the LRQ interleaved shaping algorithm and its modeling are first introduced, followed by some other preliminaries. They include traffic and server models that are used in the analysis and/or comparison. Our focused server model is GR. In Section \ref{sec-3}, the focus is on properties of interleaved LRQ. In Section \ref{sec-4}, properties of per-flow LRQ are investigated. 
In Section \ref{sec-5}, the node structure suggested by the universal approach is introduced. Following that, an improved delay bound for the current e2e delay approach is presented, with strict priority as an example scheduling discipline to discuss the improvement. Then, to demonstrate how the derived properties may be exploited, two new approaches with their e2e delay bounds are presented. 
Moreover, a discussion comparing the three approaches and their delay bounds is also provided in Section \ref{sec-5}. Finally concluding remarks are given in Sec.~\ref{sec-6}.

\section{The LRQ Algorithm and Preliminaries}\label{sec-2} 

\subsection{Notation} 

We consider FIFO systems serving flows that belong to the same class. 
A flow is a sequence of packets. Each system has one or multiple flows as inputs and outputs. In the case of multiple flows, we sometimes treat these flows as one aggregate flow.  By convention, a packet is said to have arrived to a system at the input (respectively departed the system at the output) when and only when its last bit has arrived to (respectively departed) the system. If multiple packets arrive at the same time, their original order, if it exits, is preserved; otherwise, the tie is broken arbitrarily. When a packet arrives seeing the system busy, the packet will be queued and the buffer size for the queue is assumed to be large enough ensuring no packet loss. All queues are FIFO and are initially empty. 

For a system, let $\mathcal{F}$ denote the set of flows. For each flow $f$, let $p^{f, i}$ denote the $i$-th packet in the sequence, where $i \in \mathcal{N}^+ \equiv \{1, 2, \dots \}$, and $l^{f, max}$ its maximum packet length. For every packet $p^{f, i}$, we denote by $a^{f,i}$ its arrival time to the system, $d^{f,i}$ its departure time from the system, and $l^{f,i}$ its length. The maximum packet length of the system is denoted by $l^{max}$. 
In addition, we use $A^f(s,t)$ and $D^{f}(s,t)$ to respectively denote the amount of traffic of flow $f$ which arrives to and departs from the system within time period $(s, t]$, with $A^f(t) \equiv A^f(0,t)$ and $D^{f}(t) \equiv D^{f}(0,t)$ and $A^f(0)=0$ and $D^f(0)=0$.  

Sometimes, reference time functions are used to characterize how the flow $f$ is treated by the system. Specifically, we use $E^f$ and $F^f$ to respectively refer to the times when the packets have reached the head of the queue and become eligible for receiving service, and the times when packets are expected to depart. They define reference eligible time $E^{f, i}$ and expected finish time $F^{f, i}$ for each packet $p^{f,i}$ of the flow $f$. 

For a composite system consisting of multiple systems, subscripts will be added. For instance, for a node $n$ in a network, and for a link $m$ at the node $n$, $x_n$ and $x_{n,m}$ will be respectively used, where $x$ may be any of the parameters introduced above. 

As a summary, the notation uses the form of $x_{n,m}^{f,i}$, where $n$ represents a system, e.g. a node, and $m$ a subsystem in $n$, e.g., a link; $f$ represents a flow, and $i$ the packet number in the flow. $x$ may be a packet level parameter, e.g., packet arrival time $a$, departure time $d$, length $l$; a flow level parameter, e.g.,  rate $\rho$ and  burstiness $\sigma$ for the traffic constraint and reserved or guaranteed service rate $r$, and $A$ and $D$ for cumulative traffic amount; a reference time, e.g., $E$ for eligibility time / virtual start time and $F$ for virtual finish time. When it is clear from the context, some of the superscripts or subscripts may be omitted.

\subsection{The LRQ Interleaved Shaping Algorithm}
\subsubsection{The LRQ algorithm} Length Rate Quotient (LRQ) is the first algorithm of interleaved shaping~\cite{ECRTS16}.  
Consider an LRQ shaper, whose FIFO queue is shared by an aggregate of flows. The LRQ shaper performs per-flow {\em interleaved shaping} on the aggregate, according to the algorithm shown in Algorithm~\ref{alg-lrq}~\cite{ECRTS16}. 
 
 \begin{algorithm}[ht!]
\caption{Pseudo code of the LRQ algorithm}  
\textbf{Initialization:} $\forall f: [f].eligibility\_time = 0$  \\
\textbf{Shaping:} 
\begin{algorithmic}[1]

\State {\bf while} (true) \{  
     \State \quad {\bf wait until} $q.size >0$; 
      \State   \quad     $p:= q.head(); l:=p.length; f:=p.flow\_index; $
      \State   \quad     $E^f := [f].eligibility\_time$; 
\State 
      \State \quad {\bf wait until} $t^{now} \ge E^f$; {\bf output} $p$;
\State 
     \State  \quad $E^f := t^{now} + \frac{l}{r^f}$; 
     \State \quad $[f].eligibility\_time := E^f$;
\State  \}
 
\end{algorithmic}
\label{alg-lrq}
\end{algorithm}

The LRQ algorithm shown in Algorithm~\ref{alg-lrq} takes the original form in ~\cite{ECRTS16}. As is clear from Algorithm~\ref{alg-lrq}, there is only one FIFO queue $q$ where per-flow shaping is conducted. 
In Algorithm~\ref{alg-lrq}, $q$ denotes the queue of the shaper,  where packets join in the order of their arrival times. After reaching the head of the queue,  the head packet $p$ is checked for its eligibility of output from the queue, which depends on the flow $f$ that it belongs to. Time stamp $E^f$ stores the eligible time of flow $f$ for its next packet. At the output time $d$ of packet $p$, the time stamp $E^f$ is updated to equal the present / output time $d (=t_{now})$ plus the quotient ($l/r^f$), where $l$ is the length of $p$. In this way, the next packet of flow $f$ after this packet $p$ is at least delayed until the time $t_{now}$ reaches $E^f$.

\subsubsection{A model for LRQ} 
To model the LRQ algorithm, let $j$ denote the packet number of $p$ in Algorithm \ref{alg-lrq}, i.e., $p$ is the $j$-th packet of the aggregate flow $g$ coming out of the queue $q$ in Line 3. In addition, let $a^j$ and $d^j$ denote the arrival time and output / departure time of the packet. Furthermore, let $f_{(j)}$ denote the flow where the packet is from, $i_{(j)}$ its packet number in this flow $f_{(j)}$, and $E^{f,i}$ the eligibility time of packet $p^{f,i}$, i.e., the $i$-th packet of flow $f$. 

Line 6 tells that under the condition implied by Line 2, LRQ outputs the packet immediately when the present time $t_{now}$ reaches the eligibility time of the packet $E^{f_{(j)}, i_{(j)}}$. In other words, the output time equals the eligibility time, i.e., $d^j = E^{f_{(j)}, i_{(j)}}$. The condition of Line 2 is that the packet must have already arrived, i.e. $d^ j \ge a^j$. In addition, the loop, particularly the two highlighted lines, Lines 2 and 6 imply the FIFO order is preserved when outputting packets, or in other words, $d^ j \ge d^{j-1}$. Combining these, we have
\begin{equation} \label{is-form}
d^j = \max \{a^j, d^{j-1}, E^{f_{(j)}, i_{(j)}} \}
\end{equation}
with {\em the initialization condition} $E^{f, 1} = 0$ for $\forall f$,  and $d^0 = 0$ since the queue is initially empty, where the eligibility time function $E^f$ is updated according to Lines 8 and 9 as:
\begin{equation} \label{is-con}
E^{f_{(j)}, i_{(j)}+1} = d^j + l^j / r^{f_{(j)}}. 
\end{equation}

\subsubsection{Remark on model difference} \label{sec-rm}

The concept of interleaved shaping has been extended to consider other shaping constraints, such as token-bucket constraint ~\cite{ECRTS16} \cite{Std-Qcr} and ``Pi-regularity'' constraint \cite{LeBoudec18}, and has been adopted by IEEE TSN  \cite{Std-Qcr} and IETF DetNet \cite{DetNet}. In these standards as well as in the modeling work \cite{LeBoudec18},  the interleaved shaping algorithms directly take (\ref{is-form}) as the form, where the eligibility time function (\ref{is-con}) is adapted according to the targeted shaping constraint. Specifically, the corresponding time functions of $d$  and $E$ are respectively called $GroupEligibilityTime$ and individual flow' $schedulerEligibilityTime$ in the IEEE Standard 802.1Qcr \cite{Std-Qcr}. 

In the modeling work \cite{LeBoudec18}, the introduced $\Pi^{f}$ function is indeed the function $E^{f}$ (\ref{is-con}) here. For interleaved LRQ, the Pi-function has the following expression: 
\begin{eqnarray}
\Pi_{LRQ}^{f, i}(r^f) &=& d^{f, i-1} + l^{f, i-1} / r^{f} \quad \textrm{for} \quad i \ge 2 \nonumber \\
\Pi_{LRQ}^{f, 1}(r^f) &=& - \infty \qquad \qquad \qquad \textrm{for} \quad i = 1 \label{i-con-2}
\end{eqnarray} 

As a highlight, the initial condition for the $\Pi^{f}$ function is different from the initial condition for $E^{f}$. While it is $E^{f, 1}=0$ in the initial LRQ algorithm ~\cite{ECRTS16} and the model (\ref{is-form}) above, it is $\Pi_{LRQ}^{f, 1}=-\infty$ in \cite{LeBoudec18}. Also in \cite{LeBoudec18}, this initial condition is discussed to be necessary for its proposed ``Pi-regularity'' traffic constraint model. 

\subsection{Flow and Server Models}

\subsubsection{Flow models}
For flows, two specific traffic models are considered. One is the $g$-regularity model \cite{Chang00}, also known as the {\em max-plus arrival curve} model \cite{JJ09, JL17}:
\begin{definition}
A flow is said to be $g$-regular for some non-negative non-decreasing function $g(\cdot)$ iff for all $i \ge j \ge 0$, there holds
$$ a^{i} \ge a^{j} +  g(L(i) - L(j)) $$ or equivalently, $\forall i \ge 0$,  
\begin{eqnarray} 
a^{i} &\ge& \sup_{0 \le k \le i}\{a^{k} + g(L(i) - L(k)) \} \equiv a \overline \otimes g^{(i)} \label{eq-g-regular} 
\end{eqnarray}
where $L(i) \equiv \sum_{k=0}^{i-1}l^{k}$ with $g(0)=0$ and $L(0)=0$, 
and $\overline\otimes$ is called the max-plus convolution operator. 
\end{definition}
In the case $g(x) = \frac{x}{r}$ with a constant rate $r$, which is equivalent to $a^{i+1} \ge a^{i} + \frac{l^i}{r}$, $\forall i \ge 1$, we also say the flow is {\bf $LRQ(r)$-constrained}. 

Another traffic model that will be used is the following {\em (min-plus) arrival curve} model. 

\begin{definition}
A flow is said to have a (min-plus) arrival curve $\alpha$, which is a non-negative non-decreasing function, iff the traffic of the flow is constrained by \cite{NetCal},  
$\forall s, t \ge 0$,  
$$
A(s, s+t) \le \alpha(t)
$$
or equivalently, $\forall t \ge 0$,
\begin{eqnarray} 
A(t) &\le& \inf_{0 \le s \le t}\{A(s) + \alpha(t-s) \} \equiv A \otimes \alpha(t) \label{eq-ac} 
\end{eqnarray}
where define $\alpha(0)=0$ and $\otimes$ is the min-plus convolution operator. 
\end{definition}

A special type of arrival curve, which will often be used in the paper, has the form: $\alpha(t) = \rho \cdot t + \sigma$. In this case, we will also say that the flow is leaky-bucket or token-bucket {\bf $(\sigma, \rho)$-constrained}. The $(\sigma, \rho)$ model was first introduced by Cruz in his seminal work \cite{Cruz91ab} that triggered the development of the network calculus theory. 

It can be verified that if a flow is $LRQ(r)$-constrained, it is also $(\sigma, \rho)$-constrained with $\sigma=l^{max}$ and $\rho= r$, i.e., having a (min-plus) arrival curve
$
\alpha(t) = r t + l^{max}.
$

As shown by the two definitions, while the $g$-regularity or max-plus arrival curve model characterizes a flow based on the arrival time $a^i$, the (min-plus) arrival curve model does so based on the cumulative traffic amount function $A(t)$. In the literature, e.g., \cite{Chang00, JL17}, the relationship between the min-plus and max-plus arrival curves has been investigated. Particularly, it has been shown \cite{JL17} that they can be converted to and are dual of each other. 

As a highlight, the (min-plus) arrival curve model has a straightforward property, which, however, is notoriously hard for the max-plus counterpart. It is the superposition property. Consider an aggregate flow. If every constituent flow $f$ of the aggregate has an arrival curve $\alpha^f(t) = \rho^f t + \sigma^f$, the aggregate has an arrival curve: $\sum_{f}\alpha^f = \sum_{f}\rho^f t + \sum_f \sigma^f$. 

\subsubsection{Server models}
For server modeling, define two reference time functions $E(\cdot) $ and $F(\cdot)$ iteratively as: $\forall i \ge 1$
\begin{eqnarray} 
E^{i}(r) = \max\{a^{i}, E^{i-1} + \frac{l^{i-1}}{r} \} \label{eq-ef} \\
F^{i}(r) = \max\{a^{i}, F^{i-1} \} + \frac{l^{i}}{r} \label{eq-ff} 
\end{eqnarray}
with $E^{0} = 0$, $F^{0} = 0$, and $l^{0} = 0$ where $r$ denotes the reference service rate.  Later $E$ will also be referred to as the eligibility time or {\em virtual start time (VST)} function, and $F$ the {\em virtual finish time (VFT)} function. 

Consider a physical link of rate $r$ serving a flow that inputs at one end of the link. Observe the output at the other end and ignore the propagation delay. Then, $E^{i}$ is the time that the first bit of packet $p^{i}$ starts to exit the link, and $F^{i}$ the time that the last bit finishes departing the link. 

The following relationship between functions $E$ and $F$ can be easily verified, e.g., see \cite{Jiang03}: $\forall i \ge 1$, 
\begin{eqnarray} 
F^{i} = E^{i} + \frac{l^{i}}{r^f} \label{eq-ef-ff} 
\end{eqnarray}

These functions, VST and VFT, have been used as basis in designing scheduling algorithms and in modeling the service provided by a system. The designed scheduling algorithms include Virtual Clock using $F$ \cite{Zhang00} and Start-time Fair Queueing using $E$ \cite{GVC97}. The models include Guaranteed Rate (GR) server \cite{GLV95} and its generalized version \cite{GV97}, and Start-Time (ST) server \cite{Jiang03}, which are respectively based on VFT and VST. 

In this paper, the Guaranteed Rate (GR) server model is adopted.  
\begin{definition}
A system is said to be a Guaranteed Rate (GR) server with guaranteed rate $r$ and error term $e$ to a flow, written as $GR(r, e)$, iff it guarantees that for any packet $p^{i}$ of the flow, its departure time satisfies \cite{GLV95, GV97}:
\begin{equation}
d^{i} \le F^{i}(r) + e \label{def-gr} 
\end{equation}
or equivalently
\begin{equation}
d^{i} \le a \overline\otimes g^{(i)} + \frac{l^{i}}{r} \label{max-gr}
\end{equation}
with $g^{(x)} = \frac{x}{r} + e$, where $\overline\otimes$ is the max-plus convolution operator.
\end{definition}

It has been shown that a wide range of scheduling algorithms, including priority, weighted fair queueing and its various variations, round robin and its variations, hierarchical fair queueing, Earliest Due Date (EDD) and rate-controlled scheduling disciplines (RCSDs), can be modeled using GR \cite{GLV95, GV97}.  For this reason and to simplify the representation, instead of presenting results for schedulers implementing specific scheduling algorithms, we use the GR model to represent them. A summary of the corresponding GR parameters of various scheduling algorithms can be found, e.g., in \cite{Jiang03}.

Considering the relationship (\ref{eq-ef-ff}), a server model may similarly be defined based on $E$, which is called the Start-Time (ST) server model, written as  $ST(r, \tau)$,  iff for any packet $p^{i}$ of the flow, the system guarantees its departure time \cite{Jiang03}: 
\begin{equation}
d^{i} \le E^{i}(r) + \tau \label{def-st}
\end{equation}
or equivalently
\begin{equation}
d^{i} \le a \overline\otimes g^{(i)} \label{max-ss}
\end{equation}
with $g{(x)} = \frac{x}{r} + \tau$ and $\tau = e + l^{max}/r$. 

As indicated by the max-plus convolution operator used in (\ref{max-gr}) and  (\ref{max-ss}), these models are server models for the max-plus part of network calculus \cite{Chang00}. In the min-plus part of network calculus, the {\em (min-plus) service curve} model is well-known. The latency-rate type  (min-plus) service curve is defined as follows.

\begin{definition}
A system is said to offer to a flow a latency-rate service curve $\beta(t) = r (t-\tau)^{+}$ iff for all $t \ge 0$ \cite{NetCal},  
\begin{eqnarray} 
D(t) &\ge& A \otimes \beta(t) \label{eq-sc} 
\end{eqnarray}
where $(x)^{+} \equiv \max\{x, 0\}$. 
\end{definition}

In \cite{NetCal, Jiang03}, the relationship between the GR model, the ST model, the latency-rate server model and the (min-plus) latency-rate service curve has been investigated. Particularly, it is shown \cite{Jiang03} that the latency-rate server model is equivalent to the start-time (ST) server model. With the relation (\ref{eq-ef-ff}), it can be verified that if a system is a $GR(r, e)$ server to a flow, it is also a $ST(r , e+\frac{l^{max}}{r})$ server and provides a latency-rate service curve $\beta$ to the flow \cite{NetCal, Jiang03}:
\begin{equation}\label{lr-ss}
\beta(t) = r [t - (e+\frac{l^{max}}{r})]^{+}
\end{equation}
Conversely, if the system is an $ST(r, \tau)$ or latency-rate server with the same parameters to the flow, it is also a $GR(r, \tau - \frac{l^{min}}{r})$ to the flow \cite{Jiang03}. 

\subsubsection{Delay and backlog bounds}\label{sec-gr-db}
With the flow and server models introduced above, the following delay and backlog bounds can be found or proved from literature results, e.g., \cite{GV97, NetCal}.

\begin{proposition}\label{pp-1}
Consider a flow served by a system. The flow has an arrival curve $\alpha$, and the system is a $GR(r, e)$ server to the flow. If $\lim_{t \to \infty}\frac{\alpha(t)}{t} \le r $, the delay of any packet $i$, i.e., $d^{i} - a^{i}$, is upper-bounded by, $\forall i \ge 1$, 
$$
d^{i} - a^{i} \le \frac{\sup_{t\ge0} [\alpha(t) - r t] }{r} + e
$$
and the backlog of the system at any time, i.e., $D(t) - A(t)$, is upper-bounded by, $\forall t \ge 0$,  
$$
D(t) - A(t) \le \sup_{t\ge0} [ \alpha(t) - r (t - e - \frac{l^{max}}{r})^{+} ] 
$$
\end{proposition}

As a special case, the flow is $(\sigma, \rho)$-constrained, i.e. $\alpha(t) = \rho t + \sigma$. If $\rho \le r$, the bounds in Proposition \ref{pp-1} can be written more explicitly as,  $\forall i \ge 1$, 
\begin{equation}
d^{i} - a^{i} \le \frac{\sigma}{r}+e
\end{equation}
for delay and $\forall t \ge 0$,  
\begin{equation}
D(t) - A(t) \le \sigma + \rho \cdot (e + \frac{l^{max}}{r})
\end{equation}
for backlog. 

In the TSN / DetNet literature, the delay and backlog bounds are derived commonly based on the assumption that the flow has a (min-plus) arrival curve and the server has a latency-rate (min-plus) service curve \cite{Zhao21}, except in the initial interleaved shaping paper \cite{ECRTS16} that adopts a timing analysis technique directly on the reference time functions similar to our analysis in this paper. It has also been noticed that the delay bounds from the service curve analysis are more pessimistic than from the timing based analysis \cite{Zhao21}. This difference is also seen here as discussed in the following.

Specifically, service curve-based analysis can result in a delay bound that is $\frac{l^{max}}{r}$ larger than the bound from GR-based analysis shown in Proposition \ref{pp-1}. The difference is due to the extra term $\frac{l^{max}}{r}$ in the service curve characterization as shown in (\ref{lr-ss}). By exploiting an advanced property of network calculus (NC), which is ``the last packetizer can be ignored for delay computation'' (see e.g. \cite{NetCal}), the packetizer delay can be deducted from the service curve based delay bound. However, considering that the delay bound must hold for all packets, only $\frac{l^{min}}{r}$ may thus be extracted. Consequently, the ``improved'' service curve based delay bound becomes:
$$
\frac{\sigma}{r} + e + \frac{l^{max}}{r} - \frac{l^{min}}{r}.
$$
Then its difference from GR-based analysis can be reduced to 
$$\frac{l^{max}}{r} - \frac{l^{min}}{r}.$$

As a remark, the discussion on the delay bound difference is only based on the server models themselves. When delay bound analysis is conducted on a specific scheduling discipline, the GR-based analysis may benefit additionally. As an example, strict priority will be considered and the bounds derived from different approaches be compared in Section \ref{sec-sp}.

\section{Properties of Interleaved LRQ Shapers} \label{sec-3}

In this section, we first review the ``shaping-for-free'' property of interleaved shaping and prove it for LRQ without altering the initial condition introduced for the original LRQ algorithm.
Then, we prove properties of interleaved LRQ shapers as stand-alone elements, including delay and backlog bounds.
In the next section, i.e., Section \ref{sec-4}, properties of per-flow LRQ, including per-flow LRQ based flow aggregation, are investigated.  

\subsection{The ``Shaping-for-Free'' Property} 

As introduced in Section \ref{sec-2}, functions (\ref{is-form}) and (\ref{is-con}) capture the essence of the LRQ algorithm. In addition, by adapting (\ref{is-con}), interleaved shaping of flows with other traffic constraints can be implemented, for which, a systematic investigation has been conducted in~\cite{LeBoudec18}. 

Applying (\ref{is-con}) to (\ref{is-form}), we can rewrite and obtain the following model for LRQ: $\forall j \ge 1$, 
\begin{eqnarray}\label{lrq-dtf}
 d^{j} &=& \max \{a^{j}, d^{j-1}, d^{f_{(j)}, i_{(j)}-1} + \frac{l^{f_{(j)}, i_{(j)}-1} }{r^{f_{(j)}}} \} 
\end{eqnarray}
with {\bf the initial condition: $d^{f,0}=0$ and $l^{f,0}=0$ for $\forall f$, which is equivalent to the initial condition $E^{f,0} = 0 $} for (\ref{is-form}), since the three involved parameters $d$, $l$ and $r$ in (\ref{is-con}) are non-negative in nature and $r$ is non-zero. 

In the literature, ``shaping-for-free'' is a well known property of per-flow shapers. Specifically,  if a shaper is greedy and the initial traffic constraint of the flow is used as the shaping curve, the worst-case delay of the flow in a system composed of the shaper and a server is not increased in comparison with a system of the server only, in spite of the order of the shaper and the server in the combined system. Earlier works include \cite{Zhang93} \cite{GGPS96} and a more systematic investigation is summarized in \cite{Chang00} and \cite{NetCal}. 

Under interleaved shaping, the shaping-for-free property is first studied in \cite{ECRTS16}. In \cite{LeBoudec18}, a generalized treatment is provided, where the property is proved for a wide range of traffic constraints, including both Chang's $g$-regularity and (min-plus) arrival curve constraints. 

\begin{figure}[th!]
\centering
  \includegraphics[width=0.75\linewidth]{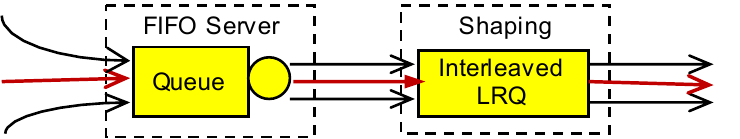} 
  \caption{The shaping-for-free property setup} 
  \label{fig-s4f}
\end{figure}

For LRQ, the shaping-for-free property is summarized in Theorem \ref{th-s4f}. Figure \ref{fig-s4f} illustrates a typical setup when studying the shaping-for-free property. 
As highlighted in Section \ref{sec-rm}, the initial condition (\ref{i-con-2}) used in \cite{LeBoudec18}, which is considered necessary there, is different from the initial condition (\ref{is-con}) used by the original LRQ algorithm \cite{ECRTS16}. In this paper, we keep the initial condition  (\ref{is-con}), which can be also written as $d^{f,0}=0$ and $l^{f,0}=0$, and re-prove the shaping-for-free property for LRQ. 
To account for the impact of the initial condition, the proof uses strong induction. 

\begin{theorem}\label{th-s4f}
Consider a set of flows $\mathcal{F}$, where every flow  $f (\in \mathcal{F})$ is $LRQ(r^f)$-regulated, i.e., $a^{f,i} \ge a^{f, i-1} + \frac{l^{f, i-1}}{r^f}$. These flows pass through a system composed of a FIFO server and an interleaved LRQ shaper with rate $r^f$ for $f$, $\forall f \in \mathcal{F}$. No matter about the order of the server and the shaper, a delay upper bound for the FIFO server is also a delay upper bound for the composite system. 
\end{theorem}

\subsection{Properties of LRQ as Stand-alone Elements}\label{sec-3-2}

In this subsection, a number of properties of LRQ as stand-alone elements are proved. Among them, while Lemma \ref{lrq-conf} and Lemma \ref{lrq-out} find similar properties of their per-flow counterparts, the other properties are unique to interleaved shaping. 

\begin{lemma}\label{lrq-conf}
(Conformance) Consider an interleaved LRQ shaper with a set of input flows $\mathcal{F}$, where for every flow $f \in \mathcal{F}$, rate $r^{f}$ is applied.  If at the input, every flow $f \in \mathcal{F}$ is $LRQ(r^{f})$-regulated, then the shaper introduces no delay, i.e., for every packet $p^{j}$, there holds $d^{j}  = a^{j}$. 
\end{lemma}

An implication of Lemma \ref{lrq-conf} is that at any time, there is at most one packet in the LRQ system from each flow. This information may be used for conformance check. For instance, from each flow, at most one packet is allowed and additional non-conformant packets are dropped. This way can prevent delaying other flows' packets if one flow is non-conformant to its $LRQ(r^f)$-constraint. 

The following output characterization result is immediately from (\ref{lrq-dtf}). 

\begin{lemma}\label{lrq-out}
(Output Characterization) Consider an interleaved LRQ shaper with a set of flows $\mathcal{F}$, where for every flow $f \in \mathcal{F}$, rate $r^{f}$ is applied. Regardless of the traffic constraint for each flow at the input, the output of the flow $f$ is constrained by $LRQ(r^f)$, i.e., $\forall i \ge 1$, 
$$
d^{f,i} \ge d^{f, i-1} + \frac{l^{f, i-1}}{r^f}.
$$ 
\end{lemma}

Having proved Lemma \ref{lrq-conf} and Lemma \ref{lrq-out}, we now focus on delay. Unfortunately, its worst-case analysis is notoriously challenging. In the rest of this section, we approach it step by step. First, the following result provides a sufficient and necessary condition for an LRQ shaper system to have bounded delay. 

\begin{lemma}\label{lrq-db-con}
(Sufficient and Necessary Condition) For an interleaved $LRQ$ system with rates $\{r^f \}$ for its flow set $\mathcal{F}$, the delay for any packet is upper-bounded, if and only if there exists a non-negative constant $\Delta(< \infty)$ such that, $\forall j \ge 1$,   
\begin{equation}\label{eq-ldbc}
d^{f_{(j)}, i_{(j)}-1} + \frac{l^{f_{(j)}, i_{(j)}-1} }{r^{f_{(j)}}} - a^{j}  \le \Delta
\end{equation}
and if the condition is satisfied, $\Delta$ is also an upper-bound on the delay. 
\end{lemma}

Note that, in Lemma \ref{lrq-db-con}, the condition does not assume how each flow is regulated at the input. If the flow is $LRQ(r^f)$-regulated at the input, applying this traffic condition together with $d^{f_{(j)}, i_{(j)}-1} = a^{f_{(j)}, i_{(j)}-1}$ from Lemma   \ref{lrq-conf} gives $d^{f_{(j)}, i_{(j)}-1} + \frac{l^{f_{(j)}, i_{(j)}-1} }{r^{f_{(j)}}} - a^{j}  \le 0$. In other words, the sufficient and necessary condition is satisfied with $\Delta=0$. This also confirms Lemma   \ref{lrq-conf} . 

When the flow is not $LRQ(r^f)$-regulated, the condition constant $\Delta$ is not as easily found. Additional  approaches are needed to help find delay bounds. For this, in Lemma \ref{lrq-g-vt}, we relate the departure time with a generalized version of the virtual start time and virtual finish time functions defined in (\ref{eq-ef}) and (\ref{eq-ff}). Specifically, their generalized counterparts are: $\forall j \ge 1$, 
\begin{eqnarray}
\tilde{E}^{j} &=& \max \{a^{j}, \tilde{E}^{j-1} + \frac{l^{j-1} }{r^{(j-1)}} \} \label{eq-ef-g} \\ 
\tilde{F}^{j} &=& \max \{a^{j}, \tilde{F}^{j-1} \} + \frac{l^{j} }{r^{{(j)}}} \label{eq-ff-g} 
\end{eqnarray}
with $\tilde{E}^{0} = \tilde{E}^{0} = l^0 = 0$ and $r^0 = \infty$,  
where, for ease of expression, we use $r^{(j)}$ to denote the rate of the flow that packet $j$ belongs to, i.e., $r^{(j)} \equiv r^{f_{(j)}}$. 

The difference between (\ref{eq-ef-g}) and (\ref{eq-ef}), and the difference between (\ref{eq-ff-g}) and (\ref{eq-ff}), are that while the rate in the function for each packet is the same in the latter, it may differ from packet to packet in the former. These generalized virtual start time and virtual finish time functions (\ref{eq-ef-g}) and (\ref{eq-ff-g}) are similarly defined in the generalized Guaranteed Rate server model \cite{GV97}. 

\begin{lemma}\label{lrq-g-vt}
(GR Characterization) Consider an interleaved LRQ shaper with a set of input flows $\mathcal{F}$, where for every flow $f \in\mathcal{F}$, rate $r^{f}$ is applied.  The departure time of any packet $p^{j}$ is bounded by: for $\forall j \ge 1$
\begin{eqnarray}
d^{j} 
&\le & \tilde{E}^{j}  = \tilde{F}^{j} - \frac{l^{j} }{r^{{(j)}}} \label{lrq-vst-1}
\end{eqnarray}
where  $\tilde{E}^{j}$ and $\tilde{F}^{j} $ are defined in (\ref{eq-ef-g}) and (\ref{eq-ff-g}) respectively. 
\end{lemma}

With Lemma \ref{lrq-g-vt}, the following corollary is immediately from the definition of the generalized GR server model, the corresponding delay bound analysis \cite{GV97} and Proposition \ref{pp-1}.

\begin{corollary}\label{cor-1}
The LRG regulator is (i) a generalized GR server with guaranteed rate $r = \min{r^f}$ and error term $e = -\min \frac{l^{f, min}}{r^f}$ and (ii) provides a service curve $\alpha(t) = \min_f r^f t$. 
(iii) If every flow is $(\sigma^f, \rho^f)$-constrained and $\sum_{f}\rho^f \le r$, then the delay of any packet $p^{j}$ is bounded by, $\forall j \ge 1$, 
\begin{eqnarray}
d^{j} - a^{j} &\le& \frac{\sum_{f} \sigma^f}{ \min_f r^f }  - \min_{f}\frac{l^{f, min}}{r^{f}} \label{eq-lrq-db0} 
\end{eqnarray}
and (iv) the backlog of the system at any time $t$ is bounded by: $\forall t \ge 0$, 
\begin{eqnarray}
D(t) - A(t) &\le& \sum_{f} \sigma^{f} + l^{max} \label{eq-lrq-bb0} 
\end{eqnarray}
\end{corollary}

While it is encouraging to have the delay bound (\ref{eq-lrq-db0}) for interleaved LRQ shapers as the first step, the condition $\sum_{f}\rho^f \le \min_f r^f$ and the term $\min_f r^f$ in  (\ref{eq-lrq-db0}) make the bound conservative. We improve in the follow result. 

\begin{theorem}\label{th-2}
Consider an interleaved $LRQ$ shaper with rates $\{r^f \}$ for its flow set $\mathcal{F}$. If every flow $f (\in\mathcal{F})$ is $(\sigma^f, \rho^f)$-constrained, and $\sum_{f} \frac{\rho^f}{r^f} \le 1$, the delay of any packet $p^{j}$ is bounded by, $\forall j \ge 1$,
\begin{equation}\label{eq-lrq-db01}
d^j - a^j \le \sum_{f}\frac{\sigma^f}{r^f}  - \frac{l^{f, j}}{r^{f}} 
\end{equation}
which implies the following delay bound for all packets:
$$
\sup_{j \ge 1} [d^j - a^j] \le \sum_{f}\frac{\sigma^f}{r^f}  - \min_{f}\frac{l^{f, min}}{r^{f}} 
$$
\end{theorem}

\section{Properties of Per-Flow LRQ} \label{sec-4}

In this section, the focus is on per-flow LRQ. We first discuss the relation of per-flow LRQ with two existing concepts / models and briefly summarize its properties corresponding to those of interleaved LRQ. Then, we investigate applying per-flow LRQ in flow aggregation. 

\subsection{Per-flow LRQ}

Unlike interleaved LRQ, whose properties were previously little investigated, much more for per-flow LRQ can be readily obtained from existing results, due to its relationship with two existing concepts, $g$-regulator \cite{Chang00} and smoothing Leaky Bucket (sLB) \cite{Jiang06-cn}. 

First, for per-flow LRQ, since there is only one flow, $f_{(j)} = f$ and packet $p^{f_{(j)}, i_{(j)}-1}$ is $p^{j-1}$. So we have
\begin{eqnarray}
 d^{j} &=& \max \{a^{j}, d^{j-1}, d^{j-1} + \frac{l^{j-1}}{r} \} 
 = \max \{a^{j}, d^{j-1} + \frac{l^{j-1}}{r} \} 
\end{eqnarray}
which, after applied iteratively, leads to
\begin{eqnarray}\label{lrq-min-g}
d^{j} = \max_{0 \le k \le j}\{a^k + g(L(j) - L(k)) \}
\end{eqnarray}
with the two functions $g(\cdot)$ and $L(\cdot)$ given as: $g(x)=\frac{x}{r}$ and $L(j) = \sum_{k=0}^{j-1}l^k$ with $g(0)=0$ and $L(0)=0$. 
Equation (\ref{lrq-min-g}) is exactly the same as how a minimal $g$-regulator is constructed (see Theorem 6.2.2, \cite{Chang00}). Hence, all related results for minimal $g$-regulators in \cite{Chang00} also apply to per-flow LRQ.

Second, there is another shaping concept equivalent to per-flow LRQ, which is smoothing Leaky Bucket (sLB) \cite{Jiang06-cn}: 
\begin{itemize}
\item[] A smoothing Leaky Bucket (sLB) is a shaper that consists of a bucket and a buffer. The bucket has two states, EMPTY and FULL, and is initially set to be EMPTY. {\em When the bucket becomes EMPTY, the sLB sends out instantaneously the head of queue packet if the buffer is not empty, and at the same time places into the bucket a number of tokens equal to the size of this packet and changes the bucket state to FULL.} The bucket leaks at a constant leaking rate. Whenever the bucket becomes empty, its state is set to be EMPTY.
\end{itemize}

Note that the key idea of LRQ is to ``hold'' the next packet till the intended time gap from the previous packet is reached. The specific mechanism of sLB can also be used to equivalently implement such holding. With this equivalence, results for sLB, e.g. in \cite{Jiang06-cn}, also carry over to per-flow LRQ. 

It is worth highlighting that an sLB shaper differs from a normal leaky bucket (LB) or token bucket (TB) even when the bucket size of LB / TB is set to be the maximum packet length \cite{Jiang06-cn}. The reason is that, sLB ensures spacing between two consecutive packets to be equal to the length rate quotient (LRQ), while LB / TB may output more than one packet at once or output packets whose spacing is closer than by sLB, unless all packets have the same length. 

Below we summary the properties of per-flow LRQ in accordance with what have been reported for interleaved LRQ. As discussed above, more results can be found following the $g$-regulator and sLB concepts, see, e.g., \cite{Chang00} \cite{Jiang06-cn}. 

\begin{corollary} (Conformance) A per-flow LRQ shaper with rate $r$ is a minimal $g$-regulator with $g(x) = \frac{x}{r}$. \end{corollary}

Since per-flow LRQ is a special case of interleaved LRQ with only one flow,  all properties discussed in the previous section also hold for per-flow LRQ. As an example,  we have Corollary \ref{lrq-out-f} and Corollary \ref{lrq-gr-f}, which will be used in later analysis, respectively from Lemma  \ref{lrq-out}  and Lemma \ref{lrq-g-vt} : 

\begin{corollary}\label{lrq-out-f}
(Output) For per-flow LRQ with rate $r^f$, the output has an arrival curve $r^f t + l^{f, max}$. In addition, if the input is $(\sigma^f, \rho^f)$-constrained with $\rho^f \le r^f$, the output is also $(\sigma^f, \rho^f)$-constrained \footnote{For interleaved LRQ, a counterpart of this is yet to be found. }, which, in combination of  the former, gives that the output has an arrival curve of $\min\{ \rho^f t + \sigma^f, r^f t + l^{f, max} \}$.
\end{corollary}

\begin{corollary}\label{lrq-gr-f}
(GR Characterization) For per-flow LRQ with rate $r^f$, the departure time of any packet $p^{j}$ is bounded by:
\begin{equation}
d^{j} \le E^j = F^{j} - \frac{l^{j}}{r^{f}}
\end{equation}
\end{corollary}

Corollary \ref{lrq-gr-f} implies that the per-flow LRQ shaper is a guaranteed rate server and has a service curve as summarized below.

\begin{corollary}\label{r2sm-f}
A per-flow LRQ shaper with rate $r$ is (i) a GR server with the same rate and error term $- \frac{l^{min}}{r}$, and provides (ii) a latency-rate service curve $\alpha(t) = r t$.
\end{corollary}

With Corollary \ref{r2sm-f}, the related results for GR and service curve models can also be applied to per-flow LRQ. Particularly, we present delay and backlog bounds for per-flow LRQ. 
As a highlight, while the backlog bound is the same as what would be found from existing GR analysis, e.g. Proposition \ref{pp-1}, or from service curve analysis \cite{NetCal}, an improved delay bound is presented in Corollary \ref{lrq-db-f}.

\begin{corollary}\label{lrq-db-f}
{(Delay and Backlog Bounds)} For a per-flow LRQ shaper with rate $r$, whose input has an arrival curve $\alpha(t) = \rho t + \sigma$, if $\rho \le r$,  then the maximum delay of any packet is upper-bounded by $ \frac{ \sigma }{r} - \frac{l^{min}}{r}$ 
and the maximum backlog of the shaper at any time is bounded by
$\sigma.$
\end{corollary}

\subsection{Aggregation based on per-flow LRQ}

In interleaved LRQ, flows are first treated in FIFO, i.e. their packets are ordered in the FIFO queue according to their arrival times, and then per-flow LRQ shaping is conducted in an interleaved manner, preserving the packet order.  For this ``FIFO-aggregation $\to$  (interleaved) per-flow shaping'' setup, as illustrated in Figure \ref{fig-s4f}, the ``shaping-for-free'' property of interleaved LRQ has been proved in the previous section. 

\begin{figure}[th!]
\centering
  \includegraphics[width=0.8\linewidth]{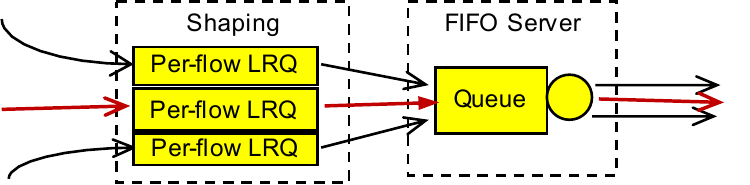} 
  \caption{LRQ-controlled aggregation} 
  \label{fig-rca}
\end{figure}

We now consider the setup ``per-flow shaping $\to$ FIFO-aggregation'', where the order of shaping and aggregation is changed. More specifically, each flow $f$ is first shaped with a per-flow $LRQ(r^f)$ shaper and the outputs from these shapers are then FIFO-aggregated based on packets' departure times from the shapers. Figure \ref{fig-rca} illustrates the setup, in contrast to the setup ``FIFO-aggregation $\to$  (interleaved) per-flow shaping'' shown in Figure \ref{fig-s4f}. 

For a system of per-flow LRQ shapers $+$ FIFO server shown in Figure \ref{fig-rca}, we have the following delay and backlog bounds.  

\begin{theorem}\label{th-rca}
{(Delay and Backlog Bounds)} Consider a set of flows $\mathcal{F}$ passing through a system, where each flow is shaped by a per-flow $LRQ(r^f)$ shaper, and the outputs from the shapers join a FIFO queue that is served by a $GR(r, e)$ server. Every flow $f \in \mathcal{F}$ is $(\sigma^f, \rho^f)$-constrained. If $\rho^f \le r^f$ for $\forall f \in \mathcal{F}$ and $\sum_f r^f \le r$, then for every packet $p^{f, j}$ of $f$, its delay is bounded by
\begin{equation}
\frac{\sigma^f}{r^f} + \frac{\sum_f l^{f, max}}{r} + e \label{c-db1}
\end{equation}
and the total backlog of all queues in the system is bounded by
\begin{equation}
\sum_f \sigma^f + \sum_f l^{f, max} + r e + l^{max} \label{c-bb1}
\end{equation}
\end{theorem}

As a comparison, for the same set of flows directly served by the FIFO $GR(r, e)$ server, the following delay and backlog bounds are from Proposition \ref{pp-1}: 
\begin{equation}
\frac{\sum_f \sigma^f}{r}  + e \label{c-db2}
\end{equation}
for delay, 
and the total backlog of all queues in the system is bounded by
\begin{equation}
\sum_f \sigma^f  + \sum_{f} \rho^{f} e + l^{max} \label{c-bb2}
\end{equation}

For backlog, it is easily seen that the backlog bound (\ref{c-bb1}) is higher, but with practical setting $r^f=\rho^f$ and $r = \sum_f r^f$, its difference from (\ref{c-bb2}) is only $\sum_f l^{f, max}$. 

For delay, the difference of (\ref{c-db1}) from (\ref{c-db2}) is $\frac{\sigma^f}{r^f} + \frac{\sum_f l^{f, max}}{r} - \frac{\sum_f \sigma^f}{r}$, which highly depends on $r^f$:  (\ref{c-db1}) may be smaller than (\ref{c-db2}) and vice versa. Note that the delay bound (\ref{c-db2}) applies universally to all flows with no difference, which may be preferred when all flows have the same delay guarantee requirements. However, when such requirements are diverse,  (\ref{c-db2}) implies that the configuration and control have to check (\ref{c-db2}) against the most stringent requirement. In contrast,  (\ref{c-db1}) implies a tuning knob, which is $r^f$, which may be set differently according to each flow's own, possibly diverse, delay requirement. 

At an immediate glance, the  ``per-flow shaping $\to$ FIFO-aggregation''  setup in Figure \ref{fig-rca} clearly requires more LRQ shapers than the ``FIFO-aggregation $\to$  (interleaved) per-flow shaping'' setup in Figure \ref{fig-s4f}. However, when both are applied to deliver bounded e2e latency in a network, as to be introduced in the next section, the total number of needed shapers may be the same.

\section{Achieving Bounded End-to-End Latency} \label{sec-5}

\subsection{Per-flow Scheduling or Aggregate Scheduling}
A central objective of TSN and DetNet is to deliver bounded end-to-end latency to flows \cite{TSN, DetNet}. With similar / related objectives, two Internet quality of service architectures, Integrated Services (IntServ) \cite{BCS94} and Differentiated Services (DiffServ) \cite{Blake98}, can be found. Their approaches to the delivery of e2e quality of service are fundamentally different. While IntServ mainly relies on per-flow scheduling to ensure isolation among flows and reserve resources along the e2e path, DiffServ only needs class-based aggregate scheduling at each node to provide service differentiation among flows. 

In per-flow scheduling, each flow has a dedicated queue through which it shares the service of a server, e.g., an output link, with other flows. An advantage of this per-flow-queue treatment is that it can effectively provide isolation among flows and subsequently deliver bounded e2e latency \cite{BCS94}. 
In contrast, in aggregate scheduling, flows of the same class typically share one queue, which shares the service of a server with queues of other classes. In delivering bounded e2e latency, per-flow scheduling is more advantageous over aggregate scheduling. This is due to that, under aggregate scheduling, the burstiness level of a flow can be significantly affected by other flows in the same aggregate due to sharing the same queue, and this influence can be cascaded. As a consequence, with FIFO aggregation, e2e delay bounds for general topology networks are only available under sometimes very restrictive utilization levels \cite{CB00, Jiang02}. 

There is a vast literature related to IntServ and DiffServ. One example is the network calculus theory, initially developed for performance guarantee analysis of IntServ and DiffServ networks \cite{Cruz91ab, Chang00, NetCal, SNetCal, JL17}, which has to date been heavily applied to such analysis of TSN and DetNet networks \cite{Zhao21}. 

Comparing with IntServ and DiffServ, TSN and DetNet also recommends class-based aggregate scheduling, however with using interleaved shaping to re-shape flows in the class aggregates. Surprisingly, interleaved shaping also has the shaping-for-free property as per-flow shaping. This has enabled an approach, including properly allocating queues at each node and reshaping flows to their initial traffic constraints, for the delivery of bounded e2e latency for TSN and DetNet \cite{ECRTS16}, and this approach is universal: its effectiveness does not dependent on network topologies. 

In this paper, a set of other properties of LRQ have been proved. This triggers the following question: Can they be used as basis to design new approaches to deliver bounded e2e latency? To this aim, two example approaches will be introduced in this section. Both keep the same way of allocating queues at nodes. 

In the remaining of this section, the node structure is first introduced in Section \ref{sec-5b}. In Section \ref{sec-5c}, an improved e2e delay bound, based on the analysis in this paper for the universal approach, is presented. Then in Sections \ref{sec-5d} and \ref{sec-5e}, two new approaches are introduced together with their e2e delay bounds. Finally, a discussion on the three approaches and their bounds is included in Section~\ref{sec-5f}.
 
\subsection{Node Structure}\label{sec-5b} 

\begin{figure}[th!]
\centering
  \includegraphics[width=0.85\linewidth]{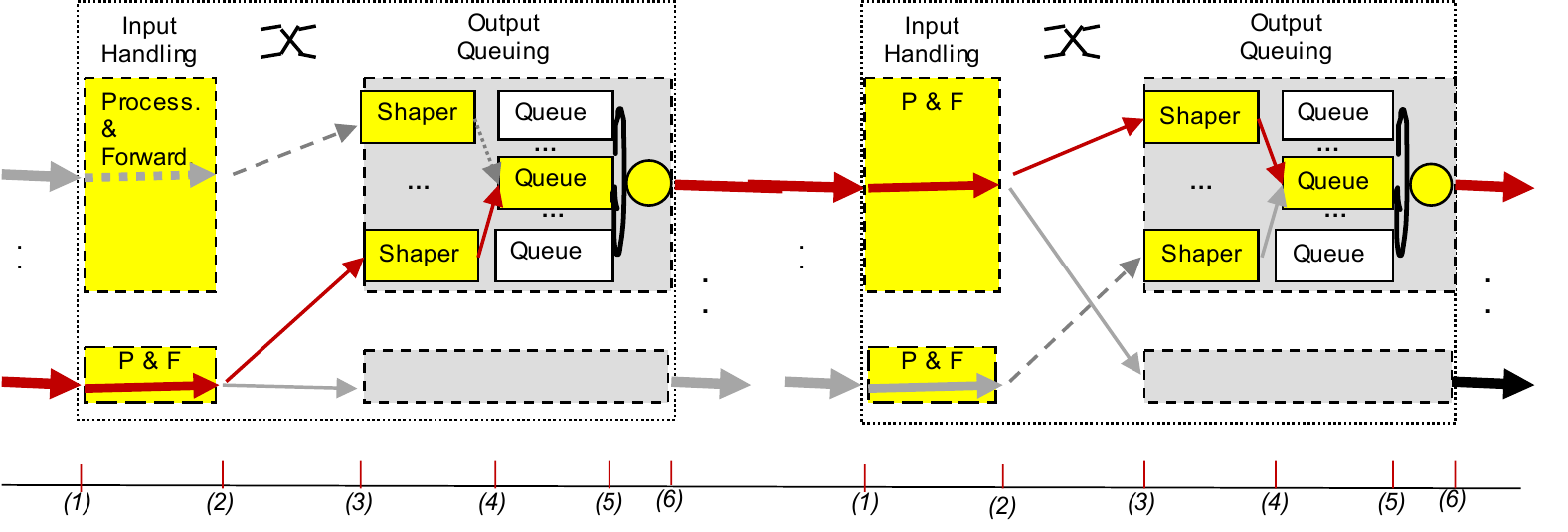} 
  \caption{Node structure and timing model} 
  \label{fig-sw}
\end{figure}

The node structure as shown in Figure~\ref{fig-sw} is adopted, which was initially proposed in \cite{ECRTS16} for TSN asynchronous traffic and has also been adopted for DetNet \cite{Finn21} to deliver bounded e2e latency. This structure was also considered earlier with the same aim but for DiffServ \cite{Jiang-net04}. 

Specifically, for an output link at the node, there are a number of queues, where each queue corresponds to one service class and is shared by its traffic in the FIFO manner. Some scheduling disciplines are employed to schedule packets from these queues in using the output link. As inputs to each queue, there is a set of shapers, where each shaper is for shaping traffic of the same class from one input link with the targeted output link \cite{ECRTS16, Jiang-net04}. 

Figure~\ref{fig-sw} also illustrates six conceptual stages that a packet goes through at a node. Upon arrival of a packet at an input link (1), it is processed and forwarded (2), via some internal mechanism or switch fabric (3), to the corresponding output queuing part of the output link, first going through a shaper (4) and then the class queue (5) followed by being served and transmitted on the output link (6). Between two consecutive stages, e.g., $(1) \to (2)$, there is delay to the packet. The delay between the first and the last stages, i.e., $(1) \to (6)$, is the delay of the packet at the node. In addition, between two adjacent nodes on the path of the packet, there is propagation delay from the last stage in the previous node to the first stage in the next node, i.e, $(6) \to (1)$, .

Without loss of generality, we assume that the service provided to each traffic class queue can be characterized using the Guaranteed Rate (GR) server model. In \cite{GV97, Jiang03}, the GR rate and error terms for a large number of scheduling disciplines can be found. 

\subsubsection{Additional notation} 
In this section, additional notation will be used in e2e delay analysis. We use $n$ to denote the node of the $n$-th output link on the path of the considered flow $f$, and $m_n$ an input link on node $n$. Let $\mathcal{F}_n$ denote the set of flows at the node $n$, which share the same output link of the flow, and $ \mathcal{F}_{n,m}$ denote the set of flows in $\mathcal{F}_n$ which are from input link $m$ of the node.  By definition, $ \mathcal{F}_n =  \cup_{m \in{ \mathcal{M}_n }} \mathcal{F}_{n,m}$, where $\mathcal{M}_n$ denotes the set of input links of the node $n$. Let $M_n$ denote the number of input links at node $n$, and $N$ the total number of nodes on the path of the considered flow $f$.

Let $r_n$ and $e_n$ respectively denote the guaranteed rate and error term of the GR characterization of the class-based aggregate scheduler for $(4) \to (6)$  at node $n$. In Section \ref{sec-5d} and Section \ref{sec-5e} for Approach 2 and Approach 3, when a per-flow LRQ shaper is used for link $m$ at node $n$, we additionally use $r_{n,m}$ to represent the rate of the corresponding per-flow LRQ shaper. 

Two traffic constraints are considered, namely $LRQ(\rho^f)$-regulated and $(\sigma^f, \rho^f)$-regulated. Accordingly, the delay bounds presented in this section are for two cases. In one case, all flows are initially $LRQ(\rho^f)$-regulated, while in the other case, they are initially $(\sigma^f, \rho^f)$-regulated. To simplify representation, the results for the two cases are combined: Specifically, we use the same $\sigma^f$ for the $(\sigma^f, \rho^f)$-regulated case and $\sigma^f = l^{f,max}$ for the LRQ-regulated case.

Corresponding to the six stages, we use $a^{f,i}_{n, (s)}$ to denote the arrival time of $p^{f,i}$ reaches the $(s)$-th stage, $s = 1, \dots, 6$, at node $n$. Thus, the delay of the packet at the node between two stages $(s1)$ and $(s2)$ is $a^{f,i}_{n, (s2)}-a^{f,i}_{n,(s1)}$. Let $T_{(s1)_{n1} \to (s2)_{n2}}$ denote an upper bound on the delay of any packet in the considered flow from $(s1)$ at node $n1$ to $(s2)$ at node $n2$. By this definition, an upper bound on the end to end delay of the flow can be written as: 
\begin{eqnarray} 
T_{(1)_{1} \to (6)_{N}} &=& \sum_{n=1}^{N} \sum_{s=1}^{5} T_{(s)_{n} \to (s+1)_{n}} \label{e2e-dbe}
\end{eqnarray}

In the investigation in Section \ref{sec-5c} to Section  \ref{sec-5e}, we focus on the output queuing part, i.e., $(3) \to (6)$ in Figure~\ref{fig-sw}, simply assuming constant delays and ignoring delay variations on $(6) \to (3)$. 
for all packets $\forall p^{f,i}$. 
Later in Section \ref{sec-5f}, the impact of delay variations on $(6) \to (3)$ on the obtained delay bounds as well as on the backlogs will be discussed. 

\subsection{Approach 1: Reshaping to the Initial Traffic Specification} \label{sec-5c}

This is the approach that was initially proposed in  \cite{ECRTS16}  for TSN asynchronous traffic. 
In this approach, the shaper at each node, cf. Figure~\ref{fig-sw}, is an interleaved shaper, which uses the initial traffic specification of each flow, i.e. either $LRQ(\rho)$ or $LB(\sigma^f, \rho^f)$-regulated, to reshape the flow's traffic at the node. 

With Theorem \ref{th-s4f} and Lemma \ref{lrq-conf}, 
we have $T_{(4)_{n} \to (4)_{n+1}}= T_{(4)_{n} \to (3)_{n+1}}$ for all $n = 1, \dots, 5$, and $T_{(1)_{1} \to (2)_{1}}= 0$ since the flow is already assumed to comply with the traffic constraint when entering the network. With these, (\ref{e2e-dbe}) can be re-written as:
\begin{eqnarray} 
T_{(1)_{1} \to (6)_{N}} &=& \sum_{n=1}^{N} T_{(4)_{n} \to (6)_{n}} + \sum_{n=1}^{N-1} T_{(6)_{n} \to (3)_{n+1}} \label{app1-dbe}
\end{eqnarray}
which indicates that if a nodal delay bound for $T_{(4)_{n} \to (6)_{n}}$ is found, an e2e delay bound can be readily obtained, since the second term on the right hand side is assumed to be bounded. 

Thanks to interleaved shaping, the traffic of every flow on $T_{(4)_{n} \to (6)_{n}}$ is shaped to its initial traffic constraint. Then a nodal delay bound for $T_{(4)_{n} \to (6)_{n}}$ can be immediately obtained from Proposition \ref{pp-1}, which is, 
$$
\frac{\sum_{f \in \mathcal{F}_n}\sigma^f}{r_n} + e_n
$$
for $\forall n$, if $\sum_{f \in \mathcal{F}_n}\rho^f \le {r_n}$. Applying the nodal delay bound to (\ref{app1-dbe}) gives an e2e delay bound summarized in Corollary \ref{app1-db}. 

\begin{corollary}\label{app1-db}
The maximum end-to-end latency of any packet of the considered flow $f$, denoted as $T_{e2e}^{App1} \equiv T_{(1)_{1} \to (6)_{N}} $, is upper-bounded by: 
\begin{eqnarray}
T^{App1}_{e2e}&\le& \sum_{n=1}^{N}  e_n + \sum_{n=1}^{N-1} T_{(6)_{n} \to (3)_{n+1}} + 
\sum_{n=1}^{N} \frac{\sum_{f \in \mathcal{F}_n}\sigma^f}{r_n} \nonumber
\end{eqnarray}
if $\sum_{f \in \mathcal{F}_n}\rho^f \le {r_n}$. 
\end{corollary} 

In Section \ref{sec-gr-db} when introducing the GR model, we have discussed that the obtained nodal delay bound is generally better than bounds from service curve analysis. A more concrete example is given below, where strict priority is focused, which has commonly been assumed as an algorithm for aggregate scheduling when conducting e2e delay analysis for TSN and DetNet \cite{Zhao21}.  

\subsubsection{Strict priority} \label{sec-sp}
The following result introduces how a strict priority server can be characterized by the GR model. 

\begin{lemma}\label{lm-sp}
Consider a flow $f$, which may be the aggregate flow of a traffic class, shares with other flows a work-conserving server of constant capacity. The server adopts non-preemptive strict priority when serving the packets. The capacity of the server is $c$. Every flow $f$ is $(\sigma^f, \rho^f)$-constrained. Then, the service provided by the server to flow $f$ can be characterized by $GR(r, e)$ with 
\begin{eqnarray}
r &=& c - \rho^u \nonumber\\
e &=& \frac{\sigma^u + l^{l, max}}{c-\rho^u} - \frac{l^{f,min}}{c-\rho^u}+ \frac{l^{f, min}}{c} \nonumber
\end{eqnarray}
and if $\rho^f \le r$, the delay of any packet of flow $f$ is bounded by $\frac{\sigma^f}{r} + e$, i.e., 
\begin{equation}\label{sp-db1}
 \frac{\sigma^f}{c - \rho^u} + \frac{\sigma^u + l^{l, max}}{c-\rho^u} - \frac{l^{f,min}}{c-\rho^u}+ \frac{l^{f, min}}{c}
\end{equation}
where $\rho^u = \sum_{f' \in \mathcal{F}^u} \rho^{f'}$ and $\sigma^u = \sum_{f' \in \mathcal{F}^u} \sigma^{f'}$
with $\mathcal{F}^u$ / $\mathcal{F}^l$  denotes the set of flows having higher / lower priority than the flow $f$, $l^{l, max}$ denotes the maximum packet length of flows in $\mathcal{F}^l$, and $l^{f, min}$ denotes the minimum packet length of flow $f$.  
\end{lemma}

As a comparison, in the original LRQ work, using a timing analysis method, the following bound has been found \cite{ECRTS16}: 
\begin{equation}\label{sp-db2}
\frac{\sigma^f}{c - \rho^u} +  \frac{\sigma^u + l^{l, max} }{c-\rho^u}+  \frac{l^{max} }{c}. 
\end{equation}
In addition, the following delay bound is obtained by using the (min-plus) service curve model \cite{Zhao21}\footnote{In \cite{ITC18}, there is effort to improve bound. However, the improvement does not cover the difference with (\ref{sp-db2}), since for the server, the service curve characterization remains the same as $c(t-l^{max})^{+}$ which cannot be avoided due to the service curve model.}:
\begin{equation}\label{sp-db3}
\frac{\sigma^f}{c - \rho^u} +  \frac{\sigma^u + l^{l, max} }{c-\rho^u}+  \frac{l^{max} }{c-\rho^u}.
\end{equation}

Clearly, the timing-analysis based bound (\ref{sp-db2}) is better than the service curve analysis based bound (\ref{sp-db3}). The difference is $\frac{l^{max} }{c-\rho^u} -   \frac{l^{max} }{c}$ per node. In addition, the GR analysis based bound (\ref{sp-db1}) makes further improvement of 
 $\frac{l^{max} - l^{f, min}}{c} + \frac{l^{f,min}}{c-\rho^u}$ over (\ref{sp-db2}). For e2e delay, the differences are multiplied, so the e2e delay bound shown in Corollary \ref{app1-db} with nodal bound (\ref{sp-db1}) may be more preferred.

\subsection{Approach 2: Reshaping to LRQ-Regularity} \label{sec-5d} 

In this and the next subsections, we investigate new approaches, exploiting the derived properties, to achieve bounded e2e latency. 
Note that, Approach 1 is universal, independent of the network topology. In this and the next subsections, we additionally take network topology information in the approach design. A simple setup is considered, where the network has a tree-topology, e.g., an aggregation network, and the focus is on flows with directions from leaves to the root. In this setup, traffic from child nodes is aggregated at their parent node that further forwards the aggregated traffic upwards to its parent node. There is no traffic segregation. 

For this setup, the total number of nodes $N$ on the path of a flow is simply the number of node generations from the entrance node of the flow to the root node, and the number of links at a node $n$, denoted as $M_n$, is the number of child nodes of $n$. 

Approach 2 is similar to Approach 1, but has three differences. One is that while all shapers in other nodes are still interleaved shapers, they are per-flow shapers in ingress nodes. Another is that, all flows from the same {\em ingress} input link and belonging to the same traffic class are treated as one (aggregate) flow $g$ and interleaved shaping is applied to such (aggregate) flows in the rest part of the network. 
The third is that all shapers are based on $LRQ$, even for the case where the initial traffic constraints are in the form of $(\sigma, \rho)$. 

Essentially, an aggregate flow $g$ represents the aggregate of flows sharing the same leave-to-root path. 
As a remark, if each flow $f$ is already treated as the e2e path-sharing aggregate flow $g$ in Approach 1, Approach 2 is the same as its LRQ-regulated version, and the only difference is that when the traffic constraint is changed to $(\sigma, \rho)$, Approach 2 still uses $LRQ$ while Approach 1 uses interleaved token bucket shapers, e.g., TBE shapers \cite{ECRTS16}. 

Without loss of generality, for a considered flow $f$ at the first node, we call its input link as {\em the first link} of the node, and denote by $r_{1,1}$ the rate of the corresponding per-flow LRQ shaper. To simplify the expression, we also denote by $r^g$ the constraint rate of aggregate $g$ which is used by the interleaved LRQ shapers. Note that for the aggregate $g$ with per-flow shaping rate $r_{1,1}$, $r^g = r_{1,1}$. 

With Lemma \ref{lrq-out}, we know the output from the first $LRQ$ shaper is $LRQ(r_{1,1})$-regulated. Thus, using interleaved shapers with the same rates $r_{1,1}$ for shapers at later nodes will not affect the corresponding delay bounds. Specifically, based on Theorem \ref{th-s4f} and Lemma \ref{lrq-conf} for $LRQ$-regulated traffic, we have $T_{(4)_{n} \to (4)_{n+1}}= T_{(4)_{n} \to (3)_{n+1}}$ for all $n = 1, \dots, 5$, which is the same as for Approach 1. However, we can no more ignore $T_{(1)_{1} \to (2)_{1}}$. With these and denoting $T_{(1)_{1} \to (6)_{N}} \equiv T_{e2e}^{App2}$, (\ref{e2e-dbe}) can be re-written as:
\begin{eqnarray} 
T_{e2e}^{App2} &=& T_{(1)_{1} \to (2)_{1}} + \sum_{n=1}^{N} T_{(4)_{n} \to (6)_{n}} + \sum_{n=1}^{N-1} T_{(6)_{n} \to (3)_{n+1}} \label{app2-dbe}
\end{eqnarray}

Let $\mathcal{G}_{n,m}$ denote the set of flows in the aggregate from the $m$-th link at node $n$, and $\mathcal{G}_{n}$ denote the set of such aggregate flows  at node $n$. 

With the delay bounds introduced in Proposition \ref{pp-1} for $(4)_{n} \to (6)_{n}$ and in Theorem \ref{th-2} for $(1)_{1} \to (2)_{1}$, an upper bound on the e2e delay for Approach 2 can be readily obtained, which is summarized in Corollary \ref{app2-db}. 

\begin{corollary}\label{app2-db}
The maximum delay of any packet of the considered flow $f$ under Approach 2, denoted as $T_{e2e}^{App2} \equiv T_{(1)_{1} \to (6)_{N}}$ is bounded by:
\begin{eqnarray}
T_{e2e}^{App2}&\le& \sum_n e_n + \sum_{n=1}^{N-1} T_{(6)_{n} \to (3)_{n+1}} + 
\frac{ \sum_{f \in \mathcal{G}_{1,1}} \sigma^{f}}{r_{1,1}} 
+  \sum_{n = 1}^{N} \frac{\sum_{g \in \mathcal{G}_n}l^{g,max}}{r_h} \nonumber
\end{eqnarray}
if $\sum_{f \in \mathcal{F}_{1,1}}\rho^f \le {r_{1,1}}$ and $\sum_{g \in \mathcal{G}_n}r^g \le {r_n}$. 
\end{corollary}

\subsection{Approach 3: Per-Input-Link Shaping with Per-Flow LRQ} \label{sec-5e}

In Approach 2, interleaved shaping is still performed inside the network. In Approach 3, we relax this requirement such that only per-flow LRQ shapers are used, in both the ingress nodes and other nodes. Other than this difference, the same setup described for Approach 2 is adopted. 

Specifically, every shaper, cf. Figure \ref{fig-sw}, treats the flows from the corresponding input link $m$ at a node $n$ as a FIFO aggregate $g_{n, m}$ and shapes the aggregate using per-flow $LRQ$ with rate $r_{n, m}$. 


While the shaper setup in Approach 3 is simpler than that in Approach 1 or Approach 2, it finds no easy way to apply the shaping-for-free property.  As a consequence, decoupling $T_{(1)_{1} \to (6)_{N}}$ into nodal elements as in (\ref{app1-dbe}) and (\ref{app2-dbe}) does not seem to have much help in its analysis. To address this change and enable e2e bounded latency analysis, a novel approach is introduced. It relies on the reference time function $F$ of the GR model and establishes the relationship between such functions between nodes. For this, the following result is important. 

\begin{lemma}\label{lm-lba-1}
Consider a flow $g_{n,m}$ sharing a $GR(r_n, e_n)$ server $n$ with other flows, where every flow is first shaped by a per-flow $LRQ$ with a rate, e.g. $r_{n,m}$ for $g_{n,m}$, and their outputs from the shapers are then fed into a FIFO queue, forming an aggregate flow $g_n$, which is served by the $GR$ server (cf. Figure \ref{fig-sw}). If $\sum_m r_{n,m} \le r_n$, we have: 
(i) the system is a $GR(r_{n,m}, e_{n,m})$ server to the flow $g_{n,m}$ with 
$$
e_{n,m} = e+ \frac{\sum_{m=1}^{M_n} l^{g_{n,m}, max}}{r_n} - \frac{l^{g_{n,m}, min}}{r_{n,m}}  
$$ 
and (ii) 
the following relation holds for any packet $p^{g_{n,m}, k}$, whose corresponding packet in the aggregate is packet $p^{g_n,j}$:  
\begin{eqnarray}
F^{g_n,j}_{(O)}(r_n) &\le& F^{g_{n,m},k}_{(I)}(r_{n,m}) + e_{n,m} + \frac{l^{max}}{r_n} 
\end{eqnarray}
where $M_n$ denotes the number of flows $g_{n,m}$ in the aggregate $g_n$, 
$F^{g_n, j}_{(O)}$ and $F^{g_n, j}_{(I)}$ respectively denote the virtual time value of packet $p^{g_n, j}$ at the output and that of the same packet in $g_{n,m}$, i.e., $p^{g_{n,m}, k}$, at the input of the system.
\end{lemma}

The relationship between the output and input's virtual time functions $F$, which is established in Lemma \ref{lm-lba-1}, may be applied iteratively from the last node to the first node, since the input at the next node is the output of the previous node. 
In addition, if there is additional but bounded delay between the output from the previous node and the input to the next node, it can also be easily factored in. 
Furthermore, at the last node, the node is a $GR$ server to the traffic from the link of the considered flow based on Lemma  \ref{lm-lba-1}.(i). 
Combining these, together with some simplification in representation, gives the following lemma. 

\begin{lemma}\label{lbfa-gr}
For a considered flow $f$ traversing the network where Approach 3 is implemented, the network is a $GR(r^f, e^f)$ server to the flow, i.e., 
$$
d_{(6)_N}^{f,i} \le F^{f,i}_{(1)_1} + e^f + \sum_{n=1}^{N-1} T_{(6)_{n} \to (3)_{n+1}} 
$$
with $r^f=r_{1,1}$, which is the rate of the first per-flow shaper that the flow goes through in the network, and  
$$
e^f = \sum_n e_n + \sum_n \frac{{(M_n+1)} \cdot l^{max} - l^{min}}{r_n} - \frac{l^{max}}{r_{N}} 
$$
if for $\forall n$, $\sum r_{n,m} \le r_n$, where $M_n$ denotes the number of links at node $n$. 
\end{lemma}

Finally, with Lemma \ref{lbfa-gr}, an e2e delay bound for flow $f$ is found from the delay bound in Proposition \ref{pp-1}, which is summarized in Corollary \ref{app3-db}. 

\begin{corollary}\label{app3-db}
The maximum delay of any packet of the considered flow $f$ under Approach 3, denoted as $T_{e2e}^{App3} \equiv T_{(1)_{1} \to (6)_{N}}$ is bounded by:
\begin{eqnarray}
T_{e2e}^{App3} 
&\le& \sum_n e_n  + \sum_{n=1}^{N-1} T_{(6)_{n} \to (3)_{n+1}} + 
\frac{ \sum_{f \in \mathcal{F}_{1, 1}} \sigma^{f}}{r_{1,1}}  +  \sum_{n=1}^{N} \frac{{(M_n+1)} \cdot l^{max} }{r_n} 
\end{eqnarray}
if $\sum_{f \in \mathcal{F}_{1, 1}} \rho^f \le r_{1,1}$ and for $\forall n$, $\sum_{m=1}^{M_n} r_{n,m} \le r_n$. 
\end{corollary}

\subsection{Discussion} \label{sec-5f}
\nop{Remark: Use a table to summarize.} 

Table \ref{table_1} presents a comparison of the delay bounds $T_{e2e}^{App1}$, $T_{e2e}^{App2}$ and $T_{e2e}^{App3}$ that can be delivered by the three approaches, Approaches 1 -- 3. To make the comparison more direct, all packets are assumed to have unit length, and all flows have the same burstiness parameter $\sigma^f$. In the table, $|\mathcal{X}|$ denotes the size of set $\mathcal{X}$. 

For the tree network, we generally have $|\mathcal{F}_n| > |\mathcal{G}_n| > |\mathcal{M}_n|$, which respectively represent the number of flows, the number of flow aggregates and the number of links at node $n$.  
This makes the 4th term in the table for Approach 1 at least $\sigma^f$ times higher. Even though for Approach 1, the 3rd term equals zero in contrast to those for Approaches 2 and 3, it applies only to one node, and for Approaches 2 and 3, their 3rd term is in the order of $\frac{ |\mathcal{F}_n| \sigma^f}{r_n}$ under practical settings. These imply that the 4th term of Approach 1 is about $N$ times of the 3rd term of Approaches 2 and 3. Jointly, we have $T_{e2e}^{App1} \gtrapprox \max\{N, \sigma^f\} T_{e2e}^{App2} \gtrapprox \max\{N, \sigma^f\}  T_{e2e}^{App3}$. 
So, in terms of delivering tight bounded e2e latency, the preference would be Approach 3 $>$ Approach 2 $>$ Approach 1.

However, the delay bounds $T_{e2e}^{App2}$ and $T_{e2e}^{App3}$ are only for the considered tree network. In contrast, the validity of $T_{e2e}^{App1}$ makes no assumption on the network topology, making the bound universally applicable. In addition, since Approach 2 still keeps interleaved shaping inside the network, its dependence on the network topology in delivering bounded e2e latency is hence less than Approach 3 where only per-flow shaping is used. Hence, in terms of reducing dependence on the topology, the preference would be Approach 1 $>$ Approach 2 $>$ Approach 3, completely opposite the above.

Note that, the key intention of introducing Approaches 2 and 3 is to demonstrate how the newly derived properties of LRQ may be useful for the delivery of TSN/DetNet qualities of service, particularly bounded e2e latency. To this aim, they have served the purpose. 

The alert reader may have noticed that, for Approaches 1 to 3, our analysis does not include backlog bounds. The reason is, with the obtained delay bounds, some backlog bounds can be readily found from a result of Network Calculus \cite{NetCal}. For a flow served by a system, if the delay is bounded by $T$, the traffic backlog of the flow in the system at any time is bounded by $\alpha(T)$ where $\alpha$ is an arrival curve of the flow. Though this backlog bound may be conservative, it is practically affordable. Take a simple example. A constant rate flow has traffic rate $8Mbps$ and maximum packet length $1KB$, implying an arrival curve $\alpha(t)=\rho t + \sigma$ with $\rho = 8Mbps$ and $\sigma = 1KB$. Its total queuing related delay in the network is bounded by $1ms$. Then at any queue in the network, at most 2KB or two maximum length packets may be found for this flow. 

As a final remark, the delay bound analysis in this section has simply assumed constant delays between stages $(6) \to (1) \to (2) \to (3)$. When these delays have jitter, the burstiness level of a flow at stage $(3)$ may increase. Notice that, such delay variations are already counted in by the second term in $T_{e2e}^{App2}$ and $T_{e2e}^{App3}$. So, the impact will mostly be on the backlog, which is likely also practically affordable as the simple example implies. 

\begin{table*} [tb!]
\footnotesize
\caption{Delay bounds delivered by Approaches 1 -- 3}
\begin{center}
\begin{tabular}{l||c|c|c|c}
\hline Approach & Term 1 & Term 2 & Term 3 & Term 4  \\\hline \hline
Approach 1 & $\sum_{n=1}^{N} e_n$ & $\sum_{n=1}^{N-1} T_{(6)_{n} \to (3)_{n+1}}$ &  & $\sum_{n=1}^{N} \frac{ |\mathcal{F}_n| \sigma^f}{r_n} $ \\\hline 
Approach 2 & $\sum_{n=1}^{N} e_n$ & $\sum_{n=1}^{N-1} T_{(6)_{n} \to (3)_{n+1}}$ & $ \frac{ |\mathcal{G}_{1,1}| \sigma^{f}}{r_{1,1}}$ & $\sum_{n = 1}^{N} \frac{|\mathcal{G}_n|}{r_n} $ \\\hline
Approach 3 & $\sum_{n=1}^{N} e_n$ & $\sum_{n=1}^{N-1} T_{(6)_{n} \to (3)_{n+1}}$ & $\frac{ |\mathcal{F}_{1, 1}| \sigma^{f}}{r_{1,1}}$ & $ \sum_{n=1}^{N} \frac{{|\mathcal{M}_n|+1} }{r_n}$ \\\hline
\end{tabular}
\end{center}
\label{table_1}
\end{table*}

\section{Conclusion} \label{sec-6}
Though being the first algorithm of interleaved shaping, the properties of LRQ were previously little studied. As a step towards filling the gap, a set of  properties for LRQ shapers have been derived in this paper, under both interleaved and per-flow shaping settings. These properties include the shaping-for-free property that has been proved without altering the initialization condition introduced in the original LRQ algorithm, and properties for LRQ shapers as standalone elements. For per-flow LRQ, its properties as a flow aggregation mechanism are additionally investigated. 
In our analysis, the Guaranteed Rate (GR) server model has been particularly exploited. The results show that an improved e2e delay bound can be obtained with this GR-based analysis for the  reshaping-to-the-initial-traffic-constraint approach. 
In addition, to illustrate how the derived properties may be exploited, two example approaches are introduced for a tree-topology network setup to deliver bounded e2e latency in the network. 
These shed new insights on the delivery of TSN / DetNet qualities of service.

\bibliographystyle{unsrt}
\bibliography{nc-qt}

\begin{thebibliography}{10}

\bibitem{ECRTS16}
Johannes Specht and Soheil Samii.
\newblock Urgency-based scheduler for time-sensitive switched ethernet
  networks.
\newblock In {\em 28th Euromicro Conference on Real-Time Systems}, 2016.

\bibitem{TSN}
IEEE.
\newblock 802.1q -- ieee standard for local and metropolitan area networks --
  bridges and bridged networks.
\newblock {\em IEEE Standards}, 2018.

\bibitem{DetNet}
N.~Finn, P.~Thubert, B.~Varga, and J.~Farkas.
\newblock Deterministic networking architecture.
\newblock {\em IETF RFC 8655}, Oct 2019.

\bibitem{Std-Qcr}
IEEE.
\newblock {IEEE} standard for local and metropolitan area networks--bridges and
  bridged networks - amendment 34:asynchronous traffic shaping.
\newblock {\em IEEE Std 802.1Qcr-2020}, pages 1--151, 2020.

\bibitem{Finn21}
N.~Finn, J-Y. {Le Boudec}, E.~Mohammadpour, J.~Zhang, B.~Varga, and J.~Farkas.
\newblock Detnet bounded latency.
\newblock {\em IETF Internet Draft: draft-ietf-detnet-bounded-latency-05},
  April 2021.

\bibitem{LeBoudec18}
J.-Y. {Le Boudec}.
\newblock A theory of traffic regulators for deterministic networks with
  application to interleaved regulators.
\newblock {\em IEEE/ACM Transactions on Networking}, 26(6):2721--2733, 2018.

\bibitem{Chang00}
C.-S. Chang.
\newblock {\em Performance Guarantees in Communication Networks}.
\newblock Springer-Verlag, 2000.

\bibitem{NetCal}
J.-Y. {Le Boudec} and P.~Thiran.
\newblock {\em Network Calculus: A Theory of Deterministic Queueing Systems for
  the Internet}.
\newblock Springer-Verlag, 2001.

\bibitem{GLV95}
P.~Goyal, S.~S. Lam, and H.~M. Vin.
\newblock Determining end-to-end delay bounds in heterogeneous networks.
\newblock In {\em Proc. Workshop on Network and Operating System Support for
  Digital Audio and Video (NOSSDAV'95)}, pages 287--298, Apr. 1995.

\bibitem{GV97}
P.~Goyal and H.~M. Vin.
\newblock Generalized guaranteed rate scheduling algorithms: A framework.
\newblock {\em IEEE/ACM Trans. Networking}, 5(4):561--571, Aug. 1997.

\bibitem{Jiang-ITC13}
Yuming Jiang.
\newblock Stochastic service curve and delay bound analysis: A single node
  case.
\newblock In {\em Proceedings of the 2013 25th International Teletraffic
  Congress (ITC)}, pages 1--9, 2013.

\bibitem{ITC18}
Ehsan Mohammadpour, Eleni Stai, Maaz Mohiuddin, and Jean-Yves Le~Boudec.
\newblock Latency and backlog bounds in time-sensitive networking with credit
  based shapers and asynchronous traffic shaping.
\newblock In {\em 2018 30th International Teletraffic Congress (ITC 30)},
  volume~02, pages 1--6, 2018.

\bibitem{Zhao21}
Luxi Zhao, Paul Pop, and Sebastian Steinhorst.
\newblock Quantitative performance comparison of various traffic shapers in
  time-sensitive networking.
\newblock {\em CoRR}, abs/2103.13424, 2021.

\bibitem{JJ09}
Jing Xie and Yuming Jiang.
\newblock Stochastic network calculus models under max-plus algebra.
\newblock In {\em IEEE Global Telecommunications Conference (GLOBECOM)}, pages
  1--6, 2009.

\bibitem{JL17}
J.~{Liebeherr}.
\newblock Duality of the max-plus and min-plus network calculus.
\newblock {\em Foundations and Trends in Networking}, 11(3-4):139--282, 2017.

\bibitem{Cruz91ab}
R.~L. Cruz.
\newblock A calculus for network delay, part {I} and part {II}.
\newblock {\em IEEE Trans. Information Theory}, 37(1):114--141, Jan. 1991.

\bibitem{Jiang03}
Y.~Jiang.
\newblock Relationship between guaranteed rate server and latency rate server.
\newblock {\em Computer Networks}, 43(3):307--315, 2003.

\bibitem{Zhang00}
Lixia Zhang.
\newblock Virtual clock: a new traffic control algorithm for packet switching
  networks.
\newblock In {\em Proc. ACM SIGCOMM'90}, 1990.

\bibitem{GVC97}
P.~Goyal, H.~M. Vin, and H.~Cheng.
\newblock Start-time {F}air {Q}ueueing: A scheduling algorithm for {I}ntegrated
  {S}ervices packet switching networks.
\newblock {\em IEEE/ACM Trans. Networking}, 5(5):690--704, Oct. 1997.

\bibitem{Zhang93}
H.~Zhang and D.~Ferrari.
\newblock Rate-controlled static-priority queueing.
\newblock In {\em IEEE INFOCOM}, pages 227--236, 1993.

\bibitem{GGPS96}
L.~Georgiadis, R.~Guerin, V.~Peris, and K.N. Sivarajan.
\newblock Efficient network qos provisioning based on per node traffic shaping.
\newblock {\em IEEE/ACM Transactions on Networking}, 4(4):482--501, 1996.

\bibitem{Jiang06-cn}
Y.~Jiang.
\newblock Delay bound and packet scale rate guarantee for some expedited
  forwarding networks.
\newblock {\em Computer Networks}, 50:15--28, 2006.

\bibitem{BCS94}
R.~Braden, D.~Clark, and S.~Shenker.
\newblock Integrated services in the {I}nternet architecture: An overview.
\newblock {\em IETF RFC1633}, 1994.

\bibitem{Blake98}
S.~Blake, D.~Black, M.~Carlson, E.~Davies, Z.~Wang, and W.~Weiss.
\newblock An architecture for differentiated services.
\newblock {\em IETF RFC2475}, 1998.

\bibitem{CB00}
A.~Charny and J.-Y. {Le Boudec}.
\newblock Delay bounds in a network with aggregate scheduling.
\newblock In {\em Proc. First International Workshop of Quality of Future
  Internet Services (QOFIS'2000)}, 2000.

\bibitem{Jiang02}
Y.~Jiang.
\newblock Delay bounds for a network of {G}uaranteed {R}ate servers with {FIFO}
  aggregation.
\newblock {\em Computer Networks}, 40(6):683--694, Dec. 2002.

\bibitem{SNetCal}
Yuming Jiang and Yong Liu.
\newblock {\em Stochastic Network Calculus}.
\newblock Springer-Verlag, 2008.

\bibitem{Jiang-net04}
Y.~Jiang.
\newblock Link-based fair aggregation: A simple approach to scalable support of
  per-flow service guarantees.
\newblock In {\em IFIP Networking Conference}, 2004.

\end{thebibliography}

\appendix

\section{Proof of Theorem \ref{th-s4f}}
The property has two parts: (I) the LRQ shaper is before the FIFO server; (II) the FIFO server is followed by the LRQ shaper as illustrated in Figure \ref{fig-s4f}. 

For part (I), the proof needs Lemma \ref{lrq-conf}  and Lemma \ref{lrq-out}, which are introduced in Section \ref{sec-3-2}.  Specifically, with the former, the regulator introduces no delay. With the latter, the output from the regulator, i.e., the input to the server, is regulated with the same traffic constraint and hence the same delay bound remains. 

For part (II), the proof is as follows. Let $\hat{a}$ denote the departure from the server and hence the arrival to the regulator. Suppose $\Delta$ is a delay bound for all packets through the FIFO server, i.e., $\hat{a}^j \le a^j + \Delta$ for $\forall j \ge 1$. We prove by strong induction that for the composite system shown in Figure \ref{fig-s4f}, $\Delta$ is also a delay bound, i.e. $d^j - a^j \le \Delta$ for $\forall j \ge 1$, where $a^j$ and $d^j$ respectively denote the arrival and departure times of the $j$-th packet through the composite system. 

For the base step, consider both the 1st and the 2nd packets. By definition and the initial condition, for the 1st packet, it is obtained immediately $d^1 = \hat{a}^{1} \le a^1 + \Delta$. 
For the 2nd packet, by the LRQ model (\ref{lrq-dtf}), $d^{2} = \max \{ \hat{a}^{2}, d^{1}, d^{f_{(2)}, i_{(2)}-1} + \frac{l^{f_{(2)}, i_{(2)}-1} }{r^{f_{(2)}}} \} \le \max \{ a^{2} + \Delta, a^{1}+\Delta, d^{f_{(2)}, i_{(2)}-1} + \frac{l^{f_{(2)}, i_{(2)}-1} }{r^{f_{(2)}}} \}$. There are two cases. (i) The 2nd packet is from a different flow, which is the first packet of that flow. In this case, $d^{f_{(2)}, i_{(2)}-1} + \frac{l^{f_{(2)}, i_{(2)}-1} }{r^{f_{(2)}}} = 0$ by definition, and hence  $d^2 \le a^{2} + \Delta$ since $a^2 \ge a^1$. (ii) The 2nd packet is from the same flow. Then, $d^{2} \le  \max \{ a^{2} + \Delta, a^{1}+\Delta, d^1 + \frac{l^{f_{(2)}, i_{(2)}-1} }{r^{f_{(2)}}} \} \le  \max \{ a^{2} + \Delta, a^{1}+\Delta, a^1 + \Delta + \frac{l^{f_{(2)}, i_{(2)}-1} }{r^{f_{(2)}}} \} = \max \{ a^{2}, a^1 + \frac{l^{f_{(2)}, i_{(2)}-1} }{r^{f_{(2)}}} \} + \Delta \le a^2 + \Delta$. This completes the base step. 

For the induction, assume the theorem holds for all packets till $j-1$ with $j>2$, which implies (i) $d^{j-1} \le a^{j-1} + \Delta$. The induction assumption also implies (ii) $d^{f_{(j)}, i_{(j)}-1} \le a^{f_{(j)}, i_{(j)}-1} +\Delta$. 
Applying these to (\ref{lrq-dtf}), together with $\hat{a}^{j} \le a^{j} + \Delta$,  gives:
\begin{eqnarray}
d^{j} &=& \max \{ \hat{a}^{j}, d^{j-1}, d^{f_{(j)}, i_{(j)}-1} + \frac{l^{f_{(j)}, i_{(j)}-1} }{r^{f_{(j)}}} \}  \nonumber\\
& \le & \max \{ a^{j} + \Delta, a^{j-1}+\Delta, a^{f_{(j)}, i_{(j)}-1} +\Delta + \frac{l^{f_{(j)}, i_{(j)}-1} }{r^{f_{(j)}}} \} \nonumber \\
&=& \max \{ a^{j}, a^{f_{(j)}, i_{(j)}-1} + \frac{l^{f_{(j)}, i_{(j)}-1} }{r^{f_{(j)}}} \} + \Delta \nonumber\\
&=& a^j +\Delta \nonumber
\end{eqnarray}
where the last step is due to the $LRQ$ traffic constraint for the flow. Note that in the induction step above, we have implicitly assume that packet $j$ is not the first packet of flow $f_{(j)}$ to apply (ii). In the case that $j$ is the first packet of flow $f_{(j)}$, by definition and the initial condition, we also have $d^{j} = \max \{ \hat{a}^{j}, d^{j-1}, 0 \} \le \max \{ a^{j}+\Delta, a^{j-1} + \Delta \} \le a^{j}+\Delta$, where we have applied the induction assumption (i).
This completes the proof. 

\section{Proof of Lemma \ref{lrq-conf}}
The proof is similar to that for the second part of Theorem \ref{th-s4f}. We prove by (strong) induction. For the base case, consider the 1st packet and the 2nd packet. By definition and the initial condition, it is obtained immediately $d^1 = a^{1}$. For the 2nd packet, $d^{2} = \max \{ a^{2}, d^{1}, d^{f_{(2)}, i_{(2)}-1} + \frac{l^{f_{(2)}, i_{(2)}-1} }{r^{f_{(2)}}} \}  = \max \{ a^{2}, a^{1}, d^{f_{(2)}, i_{(2)}-1} + \frac{l^{f_{(2)}, i_{(2)}-1} }{r^{f_{(2)}}} \} = \max \{ a^{2}, d^{f_{(2)}, i_{(2)}-1} + \frac{l^{f_{(2)}, i_{(2)}-1} }{r^{f_{(2)}}} \} $. There are two cases. (i) The 2nd packet is from a different flow. In this case, $d^{f_{(2)}, i_{(2)}-1} + \frac{l^{f_{(2)}, i_{(2)}-1} }{r^{f_{(2)}}} = 0$ by definition, and hence  $d^2 = a^{2}$. (ii) The 2nd packet is from the same flow. Then, $d^{2} = \max \{ a^{2}, d^1 + \frac{l^{f_{(2)}, i_{(2)}-1} }{r^{f_{(2)}}} \} =  \max \{ a^{2}, a^1 + \frac{l^{f_{(2)}, i_{(2)}-1} }{r^{f_{(2)}}} \} = a^2$. This proves the base case. 

For the induction, assume the theorem holds for all packets till $j-1$, which implies $d^{j-1} = a^{j-1}$ and $d^{f_{(j)}, i_{(j)}-1} = a^{f_{(j)}, i_{(j)}-1} $. Applying these to (\ref{lrq-dtf}) gives:
\begin{eqnarray}
d^{j} &=& \max \{ a^{j}, d^{j-1}, d^{f_{(j)}, i_{(j)}-1} + \frac{l^{f_{(j)}, i_{(j)}-1} }{r^{f_{(j)}}} \}  \nonumber\\
& = & \max \{ a^{j}, a^{j-1}, a^{f_{(j)}, i_{(j)}-1} + \frac{l^{f_{(j)}, i_{(j)}-1} }{r^{f_{(j)}}} \} \nonumber\\
&= & \max \{ a^{j}, a^{f_{(j)}, i_{(j)}-1} + \frac{l^{f_{(j)}, i_{(j)}-1} }{r^{f_{(j)}}} \} \nonumber\\
&=& a^j \nonumber
\end{eqnarray}
which completes the proof. 

\section{Lemma \ref{lrq-db-con}}
For proving (\ref{eq-ldbc}) is a necessary condition, let's first assume the condition does not hold and then prove the conclusion does not hold consequently. Specifically, the assumption is that for some $j$, $d^{f_{(j)}, i_{(j)}-1} + \frac{l^{f_{(j)}, i_{(j)}-1} }{r^{f_{(j)}}} - a^{j} $ is not bounded. Since by definition $d^j \ge d^{f_{(j)}, i_{(j)}-1} + \frac{l^{f_{(j)}, i_{(j)}-1} }{r^{f_{(j)}}} $ and hence  $d^j - a^j \ge d^{f_{(j)}, i_{(j)}-1} + \frac{l^{f_{(j)}, i_{(j)}-1} }{r^{f_{(j)}}} - a^{j}$, so for this $j$, $d^j - a^j $ is not bounded. This completes the necessary condition part. 

For the sufficient condition part, we prove by induction that if $(\ref{eq-ldbc})$ holds for $\forall j \ge 1$, we also have  $d^j - a^j \le \Delta$ for $\forall i \ge1$, and hence it is a delay upper-bound. For the base case, $j=1$. By definition, we have $d^{1} = a^{1}$, and hence $d^{1} - a^{1} = 0 \le \Delta$. For the induction case, let's assume $\Delta$ is an upper bound for $j-1, (\forall j>1)$ and then prove it is also an upper bound for $j$. With the definition of $d^j$, we have for its delay:
\begin{eqnarray}
d^{j} - a^j &=& \max \{ a^{j}, d^{j-1}, d^{f_{(j)}, i_{(j)}-1} + \frac{l^{f_{(j)}, i_{(j)}-1} }{r^{f_{(j)}}} \} -a^j \nonumber\\
&=& \max \{0, d^{j-1}-a^j, d^{f_{(j)}, i_{(j)}-1} + \frac{l^{f_{(j)}, i_{(j)}-1} }{r^{f_{(j)}}} -a^j\} \nonumber\\
&\le& \max \{0, d^{j-1}-a^{j-1}, \Delta \} \le   \max \{0, \Delta, \Delta \} = \Delta \nonumber
\end{eqnarray}
which completes the proof. 

\section{Proof of Lemma \ref{lrq-g-vt} }
The definitions of  $\tilde{E}^{j}$ and $\tilde{E}^{j}$ imply the following relationship between them: $\forall j \ge 1$, 
\begin{equation} \tilde{F}^{j} = \tilde{E}^{j} + \frac{l^{j} }{r^{{(j)}}} \end{equation}
which can be verified with induction. For the base step, it holds because $\tilde{F}^{1} = a^1 + \frac{l^{1} }{r^{{(1)}}}$ and $\tilde{E}^{1} = a^1$. For the induction step, under the induction assumption $\tilde{F}^{j-1} = \tilde{E}^{j-1} + \frac{l^{j-1} }{r^{{(j-1)}}}$, it also holds. 

With the fact $\tilde{E}^{j-1}  \le \tilde{E}^{j-1} + \frac{l^{j-1} }{r^{(j-1)}}$, $\tilde{E}^{j}$ can also be written as:
$$\tilde{E}^{j} = max \{a^{j}, \tilde{E}^{j-1} , \tilde{E}^{j-1} + \frac{l^{j-1} }{r^{(j-1)}} \} $$
Compare $\tilde{E}^{j}$ and $d^{j}$ that is copied below
$$d^{j} = \max \{ a^{j}, d^{j-1}, d^{f_{(j)}, i_{(j)}-1} + \frac{l^{f_{(j)}, i_{(j)}-1} }{r^{f_{(j)}}} \} $$

We prove (\ref{lrq-vst-1}) by induction. For the base case $j=1$, since $d^1 = a^1$, $ \tilde{E}^{j}  = a^1$ and the initial condition, 
(\ref{lrq-vst-1}) holds, i.e., $d^{1} \le \tilde{E}^{1}$. For the induction step, we suppose  (\ref{lrq-vst-1}) holds for all packets $1, \dots, j-1$, and consider packet $j$. There are two cases. (i) Packet $p^{j-1}$ and packet $p^{j}$ belong to the same flow. In this case, $\frac{l^{j-1} }{r^{(j-1)}} = \frac{l^{f_{(j)}, i_{(j)}-1} }{r^{f_{(j)}}}$ and hence $d^{j} \le \tilde{E}^{j}$ under the induction assumption. (ii) Packet $p^{j-1}$ belongs to a different flow.  In this case, since packet $p^{j-1}$ is the immediate previous packet of $p^{j}$, due to FIFO, packet $p^{f_{(j)}, i_{(j)}-1}$ must be an earlier packet than $p^{j-1}$, implying $d^{f_{(j)}, i_{(j)}-1} \le d^{j-1}$. Let $j^* (<j-1)$ denote the packet number of $p^{f_{(j)}, i_{(j)}-1}$ in the aggregate. 
For $\tilde{E}^{j-1}$, by applying the definition of $\tilde{E}$ iteratively, we have
\begin{eqnarray}
\tilde{E}^{j-1} 
&=& \max \{a^{j-1}, a^{j-2} +  \frac{l^{j-2} }{r^{(j-2)}}, \dots, \nonumber \\
&& \tilde{E}^{f_{(j)}, i_{(j)}-1}  + \frac{l^{f_{(j)}, i_{(j)}-1} }{r^{f_{(j)}}} + \sum_{k=j^*+1}^{j-2} \frac{l^{k} }{r^{(k)}} \} \nonumber \\
&\ge& d^{f_{(j)}, i_{(j)}-1}  + \frac{l^{f_{(j)}, i_{(j)}-1} }{r^{f_{(j)}}} \nonumber
\end{eqnarray}
where for the last step, the induction assumption $d^{f_{(j)}, i_{(j)}-1} \le  \tilde{E}^{f_{(j)}, i_{(j)}-1} $ has also been applied. With the above and the induction assumption $d^{j-1} \le  \tilde{E}^{j-1} $, the three terms in $\tilde{E}^{j} $ are all not smaller than the corresponding ones in $d^{j} $. Hence  $d^{j} \le \tilde{E}^{j}$ also holds for the second case. Combining both cases, the induction step is proved, i.e.  (\ref{lrq-vst-1}) holds for $j$. 

\section{Proof of Theorem \ref{th-2}}
For any packet $p^{j}$, there exists a packet $p^{j_0}$ whose arrival starts the ``virtual busy'' period that packet $p^{j}$ is in, where for all packets that arrive in $[a^{j_0}, \tilde{F}^j]$ there holds $a^k \le \tilde{F}^{k-1}$, $\forall k = j_0+1, \dots, j$. Alternatively, the start of the period is by the latest packet with $a^{j_0} > \tilde{F}^{j_0-1}$. 

Consider a virtual reference FIFO system which has the same input sequence $a^j$ and its output is $\tilde{F}^j$. Then this period is a busy period in the virtual reference system. Note that such a ``virtual busy'' period always exists, since in one extreme case, $p^{j_0}$ is the first packet for which $a^{1} > \tilde{F}^{0}=0$ always holds, and in another extreme case, the period is started by the packet $p^{j}$ itself and in this case, $j_0=j$. 

Applying $a^k \le \tilde{F}^{k-1}$ to the definition of $\tilde{F}^j$ gives:
\begin{eqnarray}\label{eq-}
{\tilde{F}}^{j} &=& t^0 + \sum_{k=j_0}^{j} \frac{l^k}{r^{(k)}} 
= t^0 + \sum_{m=1}^{N}\frac{W_m(t^0, {\tilde{F}}^{j})}{r^m}
\end{eqnarray}
where $W_m(t^0, d^{j}) = \sum_{k=j_0}^j l^k I_{p^k \in f}$ denotes the total amount of service (in accumulated packet lengths) from flow $f$, served in $[t^0,  {\tilde{F}}^{j}]$, where the indicator function $I_{p^k \in f}$ has the value 1 when the condition $\{p^k \in f\}$, i.e. packet $p^k$ is from flow $f$, is true. 

Because of FIFO and that the virtual system is empty at $t^0_{-}$, $W_m(t^0, d^{g,j})$ is hence limited by the amount of traffic that arrives in $[t^0, a^{f_n,i}]$: $W_m(t^0,  {\tilde{F}}^{j}) \le A_m(t^0, a^{g,j})$. 

We then have, 
\begin{equation}\label{eq-2}
{\tilde{F}}^{j} \le t^0 + \sum_{f}\frac{A^f(t^0, a^{j})}{r_m}
\end{equation}

Under the condition that $\sum_{f}\frac{\rho^f}{r^f} \le 1$, we obtain:
\begin{eqnarray}\label{eq-2}
{\tilde{F}}^{j} - a^{j} &\le& \sum_{f}\frac{A^f(t^0, a^{j})}{r^f} + t_0-a^{j} 
\le \sum_{f}\frac{\rho^f(a^{j}-t^0)+\sigma^f}{r^f} - (a^{j}- t_0) 
\le \sum_{f}\frac{\sigma^f}{r^f}  \nonumber
\end{eqnarray} 
with which, the delay bound is obtained together with Lemma \ref{lrq-g-vt}, specifically (\ref{lrq-vst-1}). 

\section{Proof of Corollary \ref{lrq-out-f}}
The first part follows directly from the interleaved version. For the second part, it is known that if the input has arrival curve $\alpha^{in}$ and the system provides a service curve of $\beta$, then the output is constrained by the arrival curve $\alpha^{out} = \sup_{s \ge 0} \{\alpha^{in}(s+t) - \beta(s)\}$ (e.g., see Theorem 1.4.3 in \cite{NetCal}). Here, we have $\alpha^{in}(t) = \rho^f t + \sigma^f$ and $\beta(t) = r^f t$. Applying them proves that the output also has an arrival curve of $\rho^f t + \sigma^f$. 

\section{Proof of Theorem \ref{th-rca}}
For packet delay of a flow, it has two parts: delay at the shaper and delay at the server. Since per-flow LRQ has a service curve $r^f$, the first part is bounded by $\sigma^f / r^f$ from Corollary \ref{lrq-db-f}. In addition, from Lemma \ref{lrq-out-f}, the output from the shaper has an arrival curve of $r^{f} t + l^{f, max}$. Applying it to the GR server, the delay at the server part is bounded by $\frac{\sum_f l^{f, max}}{r} + e $ from Proposition \ref{pp-1}. Putting both together gives (\ref{c-db1}). The backlog bound (\ref{c-bb1}) can be proved similarly. 

\section{Proof of Lemma \ref{lm-sp}}
For any packet $p^{f,i}$, suppose the departure time of $p^{f,i}$, i.e. $d^{f,i}$, is within the busy period of the server which starts at $t^0$. Note that such a busy period always exists, since in the worst case, the period is only the service time period of $p^{f,i}$ and in this case, $t^0=a^{f,i}$. 

Since the server is work-conserving with constant rate $c$ and it is busy with serving between $t^0$ and $d^{f,i}$, there holds:
\begin{equation}\label{eq-2}
d^{f,i} = t^0 + \frac{\sum_{k=(i_0)}^{(i)} l^{k}}{c},
\end{equation}
where $p^{(i_0)}$ denotes the packet whose arrival starts the busy period, and $(i_0)$ its packet sequence number and $(i)$ the packet sequence number of $p^{f,i}$ at the output. 

Among packets $p^{(i_0)}, \dots, p^{(i)}$, some belong to the considered flow $f$ and the rest the other flows. Let $p^{f,i_0}$ denote the first packet from flow $f$ in the busy period. There holds $a^{f,i_0} \ge t^0$. 
Equation (\ref{eq-2}) can be re-written as:
\begin{equation}\label{eq-3a}
d^{f,i} \le t^0 + \frac{\sum_{k=i_0}^{i} l^{f,k}}{c} + \frac{D^{c}(t^0, d^{f,i})}{c},
\end{equation}
where $D^{c}(t^0, d^{f,i})$ represents the total length (in bits) of packets from the other flows served in $(t^0, d^{f,i}]$. 
Similarly, denote $D^{u}(t^0, d^{f,i})$ such traffic from higher priority flows.

Since the busy period starts at $t^0$, this implies that immediately before $t^0$, the server is idle. In other words, all packets, which arrived before $t^0$, have been served by $t^0$. So, we have $D^{u}(t^0) = A^{u}(t^0)$, $D^{f}(t^0) = A^{f}(t^0)$, and $D^{l}(t^0) = A^{l}(t^0)$. In addition, due to priority, there is at most one packet from lower priority flows in $D^{c}(t^0)$, and if there is, it must be the packet that started the busy period. Moreover, all  higher priority flows' packets, which are served before $d^{f,i}$, must have arrived by $d^{f,i}-\frac{l^{f,i}}{c}$\footnote{This is due to being non-preemptive, so among packets from higher priority flows, only those that arrived before $p^{f,i}$ enters service at $d^{f,i}-\frac{l^{f,i}}{c}$ are served before $p^{f,i}$. In the literature, what has typically been used  is $d^{f,i}$.}. So, we have $D^{u}(d^{f,i}) \le A^{u}(d^{f,i}-\frac{l^{f,i}}{c})$. Combing these, we obtain:
\begin{equation}
D^{c}(t^0, d^{f,i}) \le D^{u}(t^0, d^{f,i}) + l^{l, max} \le A^{u} (t^0, d^{f,i}-\frac{l^{f,i}}{c})+ l^{l, max}
\end{equation}
which, when applied to (\ref{eq-3a}), results in
\begin{eqnarray}
d^{f,i} &\le& t^0 + \frac{\sum_{k=i_0}^{i} l^{f,k}}{c} + \frac{A^{c}(t^0, d^{f,i}-\frac{l^{f,i}}{c})}{c} +  \frac{l^{l, max}}{c} \nonumber \\ 
&\le &  t^0 + \frac{\sum_{k=i_0}^{i} l^{f,k}}{c} + \frac{\rho^u (d^{f,i}-\frac{l^{f,min}}{c} -t^0) + \sigma^u}{c} +  \frac{l^{l, max}}{c} \nonumber 
\end{eqnarray}
Further with simple manipulation, we obtain
\begin{equation}\label{eq-6a}
d^{f,i} \le t^0 + \frac{\sum_{k=i_0}^{i} l^{f,k}}{c-\rho^u} + \frac{\sigma^u + l^{l, max} - \frac{\rho^u}{c}l^{f, min}}{c-\rho^u} 
\end{equation}

Recall the virtual time function $F$ defined in (\ref{eq-ff}), it can be verified that, for the considered packet $p^{f,i}$, we have
\begin{eqnarray}
F^{f,i}(c-\rho^u) &\ge& a^{f,i_0} + \frac{\sum_{k=i_0}^{i} l^{f,k}}{c-\rho^u} \ge t_0 + \frac{\sum_{k=i_0}^{i} l^{f,k}}{c-\rho^u} \nonumber
\end{eqnarray}
applying which to (\ref{eq-6a}) gives 
\begin{eqnarray}
d^{f,i} &\le& F^{f,i}(c-\rho^u) + \frac{\sigma^u + l^{l, max}-\frac{\rho^u}{c}l^{f, min}}{c-\rho^u} \nonumber\\
&=& F^{f,i}(c-\rho^u) + \frac{\sigma^u + l^{l, max}}{c-\rho^u} - \frac{l^{f,min}}{c-\rho^u}+ \frac{l^{f, min}}{c}\nonumber
\end{eqnarray}
and the lemma is proved with the GR definition.

\section{Proof of Lemma \ref{lm-lba-1}}
To simplify expression, the subscript $n$ is omitted in the proof. 
To help follow the proof as well as later analysis, the illustration in Figure \ref{fig-sw} is used, where $(3) \to (6)$ is the corresponding part of the system considered in the lemma.  
Let $t_{(s)}^{f,i}$ denote the time of packet $p^{f,i}$ appears at stage $(s)$. So, $t_{(3)}^{f,i}$ and $t_{(6)}^{f,i}$ are the input and output time series of the system corresponding to $(3) \to (6)$. 

By definition, we have 
\begin{eqnarray}
F^{g_m,k}_{(3)}(r_m) &=& \max \{t^{g_m,k}_{(3)}, F^{g_m,k-1}_{(3)}  \} + \frac{l^{g_m,k}}{r_m} \\
F^{g,j}_{(4)}(r) &=& \max \{ t^{g,j}_{(4)}, F^{g,j-1}_{(4)}  \} + \frac{l^{g,j}}{r} \label{tm-1} \\
F^{g,j}_{(6)}(r) &=& \max \{ t^{g,j}_{(6)}, F^{g,j-1}_{(6)}  \} + \frac{l^{g,j}}{r} \label{tm-2} 
\end{eqnarray}

From Lemma \ref{lrq-g-vt} for LRQ, we have for the output time of $p^{g_m,k}$ from the LRQ shaper, i.e., $(3)\to(4)$ in Figure \ref{fig-sw}, 
\begin{equation} \label{tm-2}
t^{g_m,k}_{(4)} \le F^{g_m,k}_{(3)} - \frac{l^{g_m,k}}{r_m}.
\end{equation}

From Lemma \ref{lrq-out}, the output of each per-flow LRQ is $LRQ(r_m)$-regulated and has an arrival curve $r_m t + l^{g_m, max}$. With this, consider period $(t^{g,v}_{(4)_{-}}, t^{g,j}_{(4)}]$, for any $1 \le v \le j$. Clearly, the traffic in this period at $(4)$ contains all packets from $p^{g,v}$ to $p^{g,j}$ whose total length is 
$
\sum_{u=v}^j l^{g, u}. 
$
Hence, we have
\begin{eqnarray}
\sum_{u=v}^j l^{g, u} &\le& A_{(4)}^g (t^{g,v}_{(4)_{-}}, t^{g,j}_{(4)}) = \sum_m A_{(4)}^{g_m} (t^{g,v}_{(4)_{-}}, t^{g,j}_{(4)}) \nonumber \\
&\le& \sum_m [r_m \cdot (t^{g,j}_{(4)} - t^{g,v}_{(4)}) + l^{g_m, max}] \nonumber \\
&\le& r \cdot (t^{g,j}_{(4)} - t^{g,v}_{(4)}) + \sum_m  l^{g_m, max}
\end{eqnarray}
It then gives 
$$
t^{g,v}_{(4)} + \frac{ \sum_{u=v}^j l^{g, u}}{r} \le t^{g,v}_{(4)}  + \frac{ \sum_m  l^{g_m, max} }{r}.
$$
Applying the above to (\ref{tm-1}) gives 
$$
F^{g,j}_{(4)} \le t^{g,j}_{(4)}  + \frac{ \sum_m  l^{g_m, max} }{r} = t^{g_m,k}_{(4)}  + \frac{ \sum_m  l^{g_m, max} }{r}
$$
because $p^{g_m,k}$ and $p^{g,j}$ are the same packet, but counted respectively in $g_m$ and $g$, 
which together with (\ref{tm-2}) gives 
 \begin{eqnarray}
F^{g,j}_{(4)} &\le& F^{g_m,k}_{(3)}  + \frac{ \sum_m  l^{g_m, max} }{r} - \frac{l^{g_m,k}}{r_m}. \label{tm-4}
\end{eqnarray}

Since the service for $(4)\to(6)$ is $GR(r,e)$, we then have from GR definition and above:
$$
 t^{g,j}_{(6)} \le  F^{g,j}_{(4)} + e \le F^{g_m,k}_{(3)}  + e+ \frac{ \sum_m  l^{g_m, max} }{r} - \frac{l^{g_m,k}}{r_m}
$$ 
which proves that the service provided to $g_m$ can be characterized using $GR$ with rate $r_m$ and error term $
e_m = e+ \frac{ \sum_m  l^{g_m, max} }{r} - \frac{l^{g_m,min}}{r_m}$. This completes the proof for the first part. 

Also since the service for $(4)\to(6)$ is $GR(r,e)$,  we have $t^{g,j}_{(6)} \le  F^{g,j}_{(4)} + e$. Applying it to (\ref{tm-2}) and then comparing with (\ref{tm-1}), the following can be verified, same as the concatenation property of GR servers \cite{GV97}:
$$
F^{g,j}_{(6)} \le F^{g,j}_{(4)}  + e + \frac{l^{max}}{r}
$$
Applying it to (\ref{tm-4}) gives
$$
F^{g,j}_{(6)} \le F^{g_m,k}_{(3)}  + \frac{ \sum_m  l^{g_m, max} }{r} - \frac{l^{g_m,k}}{r_m} + e + \frac{l^{max}}{r}
$$
which, together with $l^{g_m, k} \ge l^{g_m, min}$, completes the proof of the second part.

\end{document}